\documentstyle[pra,aps,eqsecnum]{revtex}
\begin{document}
\title{Orthogonality catastrophe and decoherence
 of a confined Bose-Einstein condensate at finite temperature}
\author{A. B. Kuklov$^1$ and Joseph L. Birman $^2$}
\address{$^1$ Department of Engineering 
Science and Physics,
The College of  Staten Island, CUNY,
     Staten Island, NY 10314}
\address{$^2$Department of Physics, City College, 
CUNY, New York, NY 10031}

\date{\today}
\maketitle

\begin{abstract}
We discuss mechanisms of decoherence 
of a confined 
Bose-Einstein condensate at finite 
temperatures under the explicit condition of
conservation of the total number of bosons
$N$ in the trap.
A criterion for the irreversible decay
of the condensate two-time correlator 
is formulated in terms of
the {\it Orthogonality Catastrophe} (OC) for the 
exact N-body eigenstates, so that no
irreversible decay occurs without the OC.  
If an infinite external bath contacts a finite 
condensate, the OC should practically always
occur as long as the bath degrees of freedom are
interacting with each other.
We find that, 
if no external bath is present
and the role of the bath is played
by the normal component, no irreversible
decay occurs. We discuss the role
of the effect of {\it level
repulsion} in eliminating the OC.
At finite temperatures, the time-correlations
of the condensate isolated from the environment
are dominated by  
reversible dephasing which results from
the thermal ensemble averaging over realizations
of the normal component.
Accordingly, the correlator 
exhibits gaussian decay 
with certain decay time
$\tau_d$ which depends on temperature
 as well as on intensity of the shot noise determined
by the statistical uncertainty in the 
number of bosons $N$ in the trap. 
\\

\noindent PACS : 03.75.Fi, 05.30.Jp 

\end{abstract}\vskip0.5 cm 

\section{introduction}

Creation of trapped atomic 
Bose-Einstein condensates \cite{BEC}
has made possible experimental study
of the fundamental concepts which so far 
were
tested mostly in "thought 
experiments". Many intriguing questions
are associated with the finiteness of
the number of bosons $N$ forming the condensate.
As discussed by Leggett and Sols \cite{LEG,SOL},
depending on the environment,
the Josephson phase of a finite capacity Josephson
junction may exhibit either a ballistic
or a diffusive spread in time.
For example, if the condensate randomly exchanges bosons
with some environment, 
phase diffusion becomes completely
irreversible \cite{SOL}.

A dephasing of
the wave function 
of the confined condensate - the so called phase diffusion --
has been discussed in Refs. \cite{WALLS,YOU,JUHA} 
in recent times. 
The nature
of this effect is closely related to the concept of 
broken gauge symmetry associated with the
formation of the condensate wave function 
$\langle \Psi (x,t)\rangle \neq 0$, 
where $\Psi (x,t)$ is the Bose
field operator, and the averaging is performed over
the initial state taken as the {\it coherent state}. 
At zero temperature, the phase
diffusion refers to the decay of $\langle \Psi (x,t)\rangle$ 
due to the 
two-body interaction \cite{WALLS,YOU,JUHA}.
This effect is purely quantum, and occurs
without any bath.
Such a decay was predicted to be reversible in time,
so that spontaneous revivals of the wave function
should occur \cite{SOL,WALLS}. 
The phase diffusion effect in an
atomic condensate which is completely separated from
the environment was
considered by Graham \cite{GRAM1,GRAM2}, and 
it has been suggested that $\langle \Psi (x,t)\rangle$
decays irreversibly due
to the particle exchange between the condensate 
and the normal component confined in the same trap. 

The experimental study of the temporal 
phase correlations has been conducted by
the JILA group \cite{JILA}.
It has been found that no
detectable decay of the phase exists on the time scale
of the experiment $\leq 100$ms. 
The question was then posed in \cite{JILA} 
why the phase correlations are
so robust despite an apparent fast relaxation of
other degrees of freedom such as, e.g.,
the relative motion of the two condensates.

In the present work, we address the question
\cite{JILA} of the phase correlations.
However, this question is not analyzed 
in the context  of the problem of decoherence
of $\langle \Psi (x,t)\rangle$. Strictly speaking,
as long as the operator $\Psi (x,t)$ describes conserved
particles, the mean $\langle \Psi (x,t)\rangle$ is not 
physically observable.  
We analyze a physically 
measurable quantity which is the 
two-time (and space) correlator

\begin{eqnarray} 
\rho ({\bf x},{\bf x}',t, t')=\langle 
\Psi^{\dagger}({\bf x},t) \Psi ({\bf x}',t')\rangle .    
\label{1}
\end{eqnarray}
\noindent
This correlator can be measured by, e.g., 
scattering of fast atoms \cite{ATOM}.
We formulate a condition
for the irreversible decay of the correlator
(\ref{1}) in terms of the {\it 
Orthogonality Catastrophe} (OC) 
for the {\it projected N-body eigenstates}
(defined below). 

We consider a situation when the
one-particle density matrix (OPDM)
$\rho ({\bf x},{\bf x}',t)=
 \rho ({\bf x},{\bf x}',t, t)$ exhibits
equilibrium off-diagonal long
range order (ODLRO) \cite{ODLRO} (see also the
discussion in Ref.\cite{PIT} on how the ODLRO
should be defined in the atomic traps).
We ask how the existence of
the ODLRO affects the temporal behavior
of the correlator (\ref{1})
as $t-t' \to +\infty$. It is worth stressing
that this question should not be 
identified with the problem of the decay of
the broken $U(1)$ symmetry state, because such
a state is not a prerequisite
for ODLRO \cite{REM1}.
This difference becomes especially important
while studying the long time correlations in
large (but finite) $N$ condensates. As was
found by Wright et al. \cite{WALLS}, the interaction
between bosons introduces some break time 
$\sim \sqrt{N}$ \cite{REM2},
so that employing the classical field
$\langle \Psi ({\bf x}, t)\rangle$ as a proper
variable obeying
the Gross-Pitaevskii equation 
at times longer than this break time
becomes 
no longer valid.
A similar situation occurs
in the description of the normal excitations,
when the mean field approach breaks down
at long times \cite{KUK,PITCR},
with the break time being $\sim \sqrt{N}$
as well. The limitations of the Gross-Pitaevskii
equation at long times have also been pointed out
by Castin and Dum \cite{CASTIN}.
If one is interested in the long time evolution, 
the correlator (\ref{1}) must 
be analyzed.

In this paper we are interested in the long 
time correlations which, however, can be quite
measurable in the current traps. 
We calculate the condensate part of the 
correlator (\ref{1}) in equilibrium
at $T\neq 0$ and show that,
as long as no external bath is present,
the time dependence 
of the correlator (\ref{1}) can be 
represented as a result of the thermal
averaging of the $non-decaying$ exponents
where each one can be viewed as representing
a specific realization of the chemical potential
(defined below) for the exact $N$-body
eigenenergies.
This averaging results in the dephasing of
the correlator. The time $\tau_d$ of this dephasing 
is determined by the thermal 
fluctuations of the normal component for fixed $N$.
In any realistic experiment, the value
of $N$
varies statistically from run to run,
and produces shot noise,
also contributing to the dephasing rate
$\tau^{-1}_d$.
The equilibrium condensate evolution
can be viewed as though 
it occurs in the frozen
environment created by the normal component.
The dephasing, then, results from
the ensemble averaging over the possible
realizations of the normal component and $N$.

In the following
we will formulate a necessary condition for the
dissipative decay of the correlator (\ref{1})
in terms of the {\it Orthogonality Catastrophe}
for the exact $N$-body eigenstates.

\section{ Orthogonality catastrophe as a necessary 
condition for the dissipative decay of the condensate 
two-time correlator at $T\neq 0$}\label{sec:2}

The correlator (\ref{1}) carries the most complete
information about confined bosons.  
The averaging $\langle ...\rangle$ in Eq.(\ref{1})
is performed over the initial state (or states). In what
follows we will assume that the system of $N$ bosons
is in the 
thermal state, so that the averaging is performed
over the thermal ensemble of the
exact $N$-body eigenstates $|m,N\rangle$
and the eigenenergies $E_m(N)$, 
where $m$ denotes a set of the quantum numbers
specifying the eigenstate with
given $N$ of the many body Hamiltonian
$H$. In other words,
if the ground state $|0,N\rangle$ corresponds
to the pure condensate (characterized
by the energy $E_0(N)$), the states
with $m\neq 0$ describe the
condensate and the normal
component in the trap, with $m$ carrying
the meaning of the set of all 
possible quantum numbers of the 
normal excitations. Should some external bath be 
present in addition to the normal 
component, providing 
(coining the terminology \cite{JILA})
the intrinsic bath, 
these numbers
characterize both the normal component and the bath.
In the thermal equilibrium 
 Eq.(\ref{1})
is reduced to $\rho ({\bf x},{\bf x}',t, t')=
\rho ({\bf x},{\bf x}',t-t', 0)$. Accordingly,
the OPDM becomes time independent
$\rho ({\bf x},{\bf x}',t)= 
\rho ({\bf x},{\bf x}',0, 0)$.
In order to simplify notation, we will set $t'=0$. 

As a matter of fact, for temperatures $T$ below
and not very close to 
the Bose-condensation temperature $T_c$, the behavior
of (\ref{1}) is dominated by the condensate part.
The condensate part $\rho_0({\bf x},{\bf x}',t,0)$
of $\rho ({\bf x},{\bf x}',t,0)$ is defined
in terms of some macroscopically 
populated eigenstate 
$\varphi_0({\bf x}) $ of the OPDM 
\cite{ODLRO,PIT}. 
In the case of the weakly interacting Bose gas,
$\rho_0({\bf x},{\bf x}',t,0)$
can be selected by employing the following 
standard representation 

\begin{eqnarray} 
\Psi ({\bf x},t)=a_0(t)\varphi_0({\bf x}) + 
\psi' ({\bf x},t),
\label{2}
\end{eqnarray}
\noindent
 where $a_0(t)$ removes one boson from the condensate,
and $\psi'$ accounts for the
non-condensed bosons. Thus, 
$\rho_0({\bf x},{\bf x}',t,0)= \varphi^*_0({\bf x})
\varphi_0({\bf x}') \langle 
a^{\dagger}_0(t) a_0(0)\rangle $. In what follows
we will omit the coordinate dependencies of 
$\rho_0({\bf x},{\bf x}',t)$, and will call
the correlator $ \rho_0(t)=
\langle a^{\dagger}_0(t) a_0(0)\rangle $
as the condensate (time) correlator.
Accordingly, the condensate OPDM becomes
$\rho_0=\rho_0(t=0)$. Apparently,
$\rho_0$ should be identified with
the mean population $\overline{N}_0$
of the condensate.

We assume that the equilibrium condensate
can be well described by a macroscopic population
of only one single-particle  
state $\varphi_0$. 
There are situations when this assumption
is not valid. These are related to
the low D geometries 
in which the condensate may exist as a quasi-condensate
\cite{SVIST}, or is thermally smeared over many
states characterized by different winding of
the phase \cite{RING}. In this paper we do not
consider such situations. Thus, our analysis can be
applied to the 3D case only. Furthermore, if there
is the spin degeneracy, the so called fragmented
condensate can be formed \cite{FRAG}. In this case,
many states can also be macroscopically populated. 
Here
we do not consider such a situation as well.   
 
Employing
the standard definitions
of the Heisenberg operators
as well as of the thermal mean,
one finds

\begin{eqnarray} 
\displaystyle \rho_0 (t)= \langle a^{\dagger}_0(t) a_0(0)\rangle=
\sum_{m,m'}p_m 
{\rm e}^{
i[E_m(N) - E_{m'}(N-1)]t}
| (a_0)_{m',m}|^2,    
\label{3}
\end{eqnarray}
\noindent
where the Boltzmann factor is 
~$ p_m= Z^{-1}(N,\beta)
\exp(-\beta E_m(N))$, and 
$Z(N,\beta)$ denotes the canonical
partition function as a function of the total number
of bosons $N$ and of the temperature $T=1/\beta$; the 
notation

\begin{eqnarray} 
 (a_0)_{m',m}=
\langle N-1, m'|a_0| m,N\rangle, 
\label{4} 
\end{eqnarray}
\noindent
has been introduced. Eqs.(\ref{3}), (\ref{4}) are exact.
Below, we will represent $\rho_0(t)$ as a cumulant
expansion which is actually the $1/N$ expansion
in the exponent, and we will find the leading
term of this expansion.

We note that several papers \cite{GIR,GARD,CASTIN}
have been devoted to 
developing an approach for treating $N$ bosons under
the explicit condition of conservation of $N$. 
In this section, we introduce a natural
phase-space for the $N$-conserving approaches. 
When $N$ is conserved, 
it is convenient to express the matrix elements (\ref{4})
in such a way that $N$ becomes a parameter. 
First, we represent the
exact eigenstate ~$|m,N\rangle$~ of $N$ bosons as an
expansion

\begin{eqnarray} 
\displaystyle |m,N\rangle = \sum_{N_1,N_2,...}
C_{N_1,N_2...}(m,N)
|N_0,N_1,N_2,...\rangle, 
\quad N_0=N-(N_1+N_2+...), 
\label{16}  
\end{eqnarray}
\noindent
in the Fock space ~$|N_0,N_1,N_2,...\rangle$~ 
of the population numbers
~$N_0,\, N_1,\, N_2, ...$~ of some set of the single particle
states ~$\varphi_0({\bf x}),\,\varphi_1({\bf x}),\,
\varphi_2({\bf x})...$, respectively, given that
~$\varphi_0({\bf x})$~ is the only macroscopically 
populated state.
Eq.(\ref{16}) takes
into account explicitly
that the total number of bosons $N$ is
conserved; ~$C_{N_1,N_2...}(m,N)$~
denotes the coefficients of the expansion.
These coefficients 
form the Fock 
representation of the states with given $N$.
We call ~$C_{N_1,N_2...}(m,N)$~
the {\it projected states}, and introduce
a short notation ~$|\widehat{m,N\rangle }$~ 
for them.
Accordingly, the product
of two projected states is defined as  
~$\widehat{\langle N', m'}|\widehat{m,N\rangle }
=\sum_{N_1,N_2,...}
C^*_{N_1,N_2...}(m',N')C_{N_1,N_2...}(m,N)$.
Thus, for the same $N$, the product of any two
{\it projected states}  
coincides with the product
of the corresponding eigenstates (\ref{16}),
that is, ~$\widehat{\langle N, m'}|\widehat{m,N\rangle }
=\langle N, m'|m,N\rangle $. 
Therefore, the orthogonality
condition
~$\widehat{\langle N, m' } \widehat{ |m, N\rangle }
=\delta_{m'm}$ holds. 

We note that,
in general, no orthogonality exists between
two {\it projected states} with different $N$. In
other words, ~$\widehat{\langle 
N, m }| \widehat{ m',N'\rangle }=\sum_{N_1,N_2,...}
C^*_{N_1,N_2...}(m,N)C_{N_1,N_2...}(m',N')
\neq 0$~ for $N'\neq N$. 
This should be contrasted with the trivial
orthogonality ~$\langle
N, m |  m',N'\rangle =0$~ for $N'\neq N$ of the 
eigenstates (\ref{16}) insured by the orthogonality 
of the two Fock subspaces with different $N$. 
In order to clarify
the meaning of the projected states ~$ \widehat{ |m,N\rangle }$,
it is straightforward to recall the radial
part ~$R_{n,L}(r)$~ of the full wavefunction
 ~$R_{n,L}(r)Y_{L,L_z}(\theta, \phi)$~ of a particle
moving in a central potential. Then, conserving $N$
is analogous to conserving angular momentum $L$.
While $R_{n,L}(r)$ and $R_{n',L}(r)$ characterized
by different radial numbers $n$ are orthogonal for
given $L$, no orthogonality exists between
$R_{n,L}(r)$ and $R_{n',L'}(r)$ for $L\neq L'$.

The
eigen-problem ~$H |m,N\rangle = E_m(N) |m,N\rangle$~
is equivalent to 
finding
the coefficients ~$ C_{N_1,N_2...}(m,N)$~.
 Accordingly, 
this problem can 
equivalently be reformulated 
as ~$H(N)\widehat{|m,N\rangle}= E_m(N) 
\widehat{|m,N\rangle}$~ in terms of the
{\it projected states},  which are
the eigenstates of the corresponding 
$projected$
Hamiltonian $H(N)$ with the same eigenergies $E_m(N)$
for given $N$.
Resorting back to the analogy with the motion 
in a central potential, $H(N)$ can be mnemonically viewed
as the "radial" part of the total Hamiltonian.
The {\it projected states} take care explicitly
of the conservation of $N$. As a matter of fact,
the {\it projected states} form a basis for previously used
$N$-conserving approaches \cite{GIR,CASTIN,GARD}.
Employing them, it is possible, in principle, to
construct explicitly the {\it projected } Hamiltonian 
~$ H(N)=\sum_m
\widehat{|m, N\rangle }E_m(N)\widehat{\langle N,m|}$~
in terms of the {\it projected states}. It should be noted that,
despite being possible in principle, representing the
{\it projected states} ~$\widehat{|m,N\rangle}$~ in the 
original Fock basis by the coefficients ~$C_{N_1,N_2,...}(m,N)$~
is impractical as long as the interparticle interaction is
finite. The most productive way of constructing 
 the {\it projected states} is in terms of the pair operators
suggested in Refs. \cite{GIR,GARD}. We, however, will first
analyze general properties of the 
{\it projected states} without constructing them explicitly
(Sec.III).
For this purpose, any complete representation is suitable.
Then, we will demonstrate the meaning of the {\it projected 
states} by employing an exactly solvable model 
of the bath of non-interacting oscillators (Sec.IVA). 
The role of interactions between the oscillators
will be analyzed in sections IVB, IVC. 
We will also show (Sec.V) that the traditional hydrodynamical
approach \cite{HYDRO,REV}
provides a natural formalism for constructing
the {\it projected states}.

The matrix elements (\ref{4}) can be expressed in terms
of the overlaps of the {\it projected states} with $N$
differing by 1. Indeed,
a substitution of Eq.(\ref{16}) into Eq.(\ref{4})
yields

\begin{eqnarray}
\displaystyle (a_0)_{m',m}=\sum_{N_1,N_2,...}
\sqrt{N_0}C^*_{N_1,N_2...}(m',N-1)C_{N_1,N_2...}(m,N),
\label{30}
\end{eqnarray}
\noindent
where $N_0$ is given in Eq.(\ref{16}).
We consider a situation when the
mean $\overline{N}_0$ of the condensate
population $N_0$ for given $N$ 
is macroscopically large.
As was shown by Giorgini et al. 
in Ref.\cite{FLUC}, the relative mean square
fluctuation ~$\delta N_0/\overline{N}_0$ of $N_0$~ 
vanishes in the 
canonical ensemble for ~$N\gg 1$. Thus, in the calculation
of the matrix
elements (\ref{30}) one can perform an 
expansion with respect to 
~$\delta N_0/\overline{N}_0\ll 1$~, 
and retain only the zeroth
order term ~$\sqrt{\overline{N}_0}$,
so that within the accuracy 
~$o(\delta N_0/\overline{N}_0)$~ 
Eq.(\ref{30}) gives

\begin{eqnarray} 
 (a_0)_{m',m}=\sqrt{\overline{N}_0}\,\, \chi_{m',m},
\label{18}  
\end{eqnarray}
\noindent
where we have introduced the 
overlap of the {\it projected states}

\begin{eqnarray} 
\chi_{m',m}=\sum_{N_1,N_2,...}
C^*_{N_1,N_2...}(m',N-1)C_{N_1,N_2...}(m,N)=
\widehat{\langle N-1,m'}|\widehat{m,N\rangle }
\label{24}
\end{eqnarray}
\noindent
with $N$ differing by $1$. In what
follows we will assume that $\overline {N}_0\approx
N$.
 It is useful to mention
some obvious properties of the overlap (\ref{24}). Employing
the orthonormality condition for the {\it projected
states} with given $N$, one finds

\begin{eqnarray} 
\sum_{m''}\chi^*_{m',m''}\chi_{m,m''}=\delta_{m',m}.
\label{25}
\end{eqnarray}
In Beliaev's work \cite{BEL} it has
been shown that, for the exact ground state
$|0,N\rangle $, one can write
~$(a_0)_{m',0}=
\sqrt{{\overline N}_0}(\delta_{m',0} + o(1 /N))$.
In other words, creation or 
destruction of one boson in the condensate does not
lead to creation of excitations within the accuracy
$o(1/N)$ (no external bath was considered in
Ref.\cite{BEL}).  
Beliaev's result can be reformulated 
in terms of the overlap (\ref{24}) as
~$\chi_{0,0}=1 + o(1 /N)$.
This immediately yields Eq.(\ref{3}) as
~$\rho_0(t)=\overline{N}_0\exp [i(E_0(N) - E_0(N-1))t]
+o(1)$~ \cite{BEL} at $T=0$, which implies
no decay as long as $N$ is fixed.
 
At finite temperatures $T\neq 0$, 
Beliaev's result is traditionally extended
\cite{STAT} as follows
~$(a_0)_{m',m}\approx
\sqrt{\overline{N}_0}\,\,\delta_{m',m}$. In terms of the
overlap (\ref{24}), this is identical to

\begin{eqnarray} 
\chi_{m',m}=\delta_{m',m} + o(1/N). 
\label{9}
\end{eqnarray}
\noindent
We introduce the notation
~$\rho^{(d)}_0=\sum_mp_m|(a_0)_{m,m}|^2$.
Obviously, ~$\rho^{(d)}_0$, which plays
an important role in the following
analysis, is the diagonal part (with respect
to the excitation label) of the condensate
OPDM ~$\rho_0=\sum_{m,m'}p_m|(a_0)_{m,m'}|^2
=\overline{N}_0$. Employing Eq.(\ref{18}), it is
convenient to represent 
~$\rho^{(d)}_0$~ as 

\begin{eqnarray} 
\displaystyle 
\rho^{(d)}_0 =\overline{N}_0\, \overline{\chi}^2,\quad
\overline{\chi}=\sqrt{\sum_m 
p_m|\chi_{m,m}|^2},
\label{1000}
\end{eqnarray}
\noindent
where we have
introduced the {\it mean overlap}
$\overline{\chi}$.
A direct consequence of (\ref{9}) is

\begin{eqnarray} 
\overline{\chi}=1+o(1/N).
\label{100}
\end{eqnarray}
\noindent
The physical interpretation
of Eq.(\ref{100}) is that, while entering or exiting
the condensate, a boson does not significantly disturb the
excitations reservoir. 

Employing Eqs.(\ref{18}),(\ref{9}) in Eq.(\ref{3}),
we obtain the condensate correlator (\ref{3}) as
~$\rho_0 (t)=\rho^{(d)}_0 (t)+o(1)$, where
we have introduced the diagonal part (with respect
to the excitations) ~$\rho^{(d)}_0 (t)$~ 
of the
correlator (\ref{3}). Given the condition (\ref{9}), 
it is

\begin{eqnarray} 
\displaystyle 
\rho^{(d)}_0 (t)&=&\overline{N}_0\sum_m 
p_m{\rm e}^{ 
i\mu_m t}\, +o(1),
\nonumber
\\
\phantom{XXXXX}
\label{10}
\\ 
 \mu_m&=& E_m(N) - E_m(N-1)={\partial E_m(N)
\over \partial N} +o(1/N).    
\nonumber
\end{eqnarray}
\noindent
Following the standard definition \cite{STAT}, 
the quantity $\mu_m$ introduced in
Eq.(\ref{10}) will be called
the chemical potential of the $m$th exact
$N$-body eigenstate, so that the dephasing 
can be viewed as occurring due to 
the canonical ensemble fluctuations of the so
defined chemical potential.

Obviously,  
$\rho^{(d)}_0 (t)\approx \overline{N}_0\approx N$ 
for times short enough, so that no dephasing
takes place. 
It is worth emphasizing that,
while no decay is observed for each
particular exponent representing the $m$-th eigenstate,
the summation in Eq.(\ref{10}) will 
normally produce a dephasing at times longer than 
some dephasing time $\tau_d$.
We also mention that a
statistical uncertainty inevitably present 
in the initial value of $N$, namely: the 
 shot noise 
(from realization to realization), will result
in the dephasing as well \cite{REM3}. The evolution
represented by (\ref{10}) can be viewed as $reversible$
dephasing in a sense that no normal excitations 
are disturbed by such an evolution 
(see also in Ref.\cite{WALSER}). We, however,
note that this definition does not necessarily imply
that the evolution can always be reversed 
in time practically. 

An alternative to the situation described above
is that the mean overlap introduced
in Eq.(\ref{1000}) becomes ~$\overline{\chi}=o(1/N)$ .
It is important to emphasize that such a
situation does not conflict with 
the existence of ODLRO. 
Let us discuss this. Apparently, if 
~$\overline{\chi}=o(1/N)$, 
the diagonal part ~$\rho_0^{d}$~ (\ref{1000}), 
instead of being $\approx N$,
becomes

\begin{eqnarray} 
\displaystyle 
\rho^{(d)}_0=\overline{N}_0\,\,\overline{\chi}^2=o(1).
\label{26}
\end{eqnarray}
\noindent
On the other hand, the condensate OPDM
~$  \rho_0 =\overline{N}_0\sum_{m,m'}p_m
|\chi_{m,m'}|^2=\overline{N_0}\approx N$~ due
to the condition (\ref{25}). In fact,
the summation
here runs essentially over $m'\neq m$, that is, the
sum is collected from the amplitudes of the
processes when one boson, which enters
or exits the condensate, significantly disturbs 
the bath. Thus, it is conceivable that
the ODLRO, which can be expressed as ~$\rho_0\sim N$,
can coexist with the situation represented
by Eq.(\ref{26}).
We will refer to such a situation
as the {\it Orthogonality
Catastrophe} (OC) \cite{OC} in the 
system consisting of the Bose-Einstein condensate 
and the bath, which can be either extrinsic \cite{JILA}
or intrinsic \cite{JILA}.

We point out that, should the OC
occur, the dynamics of the correlator
(\ref{3})
would be totally determined by
$\chi_{m',m}$ with $m\neq m'$,
that is by the processes of creation and 
destruction as well as of scattering of the
normal excitations.
Consequently, the condensate correlator would exhibit a
dissipative decay as long as the excitations 
have finite life-time due to the many-body interactions.

The notion of the OC \cite{OC} was introduced 
by P.W. Anderson 
with
respect to the Fermi-edge singularity where
creation or annihilation of a single
hole dramatically changes all states
of the Fermi sea, so that the overlap
of the states (projected on the Fermi-sea )
which differ in the number of holes by 1 is essentially
zero in the thermodynamical limit. 
The OC in the bosonic system 
is defined above in terms of the mean square overlap
~$\overline{\chi}$~ of the
{\it projected states} which differ only in the total number
of bosons $N$ by 1. Thus, in this case the role
of the hole is played by 1 boson
removed from (or added to) the condensate, 
and the role of the Fermi
sea is played by the ensemble of the 
normal excitations (or by the external 
bath). 

Summarizing, if
~$\overline{\chi}\to 0$~, the
OC occurs, and the condensate exhibits the irreversible decay.
If ~$\overline{\chi}=1+ o(1/N)$~, no OC occurs, and the
condensate correlator (\ref{3}) is 
essentially described by Eq.(\ref{10}),
which, however, may exhibit a dephasing as a result
of the thermal averaging.
Below we will derive a general expression for 
~$\overline{\chi}$~
for the case when $N$ is macroscopically
large. We will also obtain an expression for
the correlator (\ref{3}) in the main $1/N$
limit.  

\section{Overlap and the spectral function of the bath}

If the total number of bosons $N$ is conserved, the $projected$
Hamiltonian $H(N)$ can be constructed. In this Hamiltonian,
$N$ plays a role of the parameter. Consequently,
the mean overlap $\overline{\chi}$ between
the {\it projected states} with different $N$
(see Eqs.(\ref{24}),
(\ref{1000})) can be found by 
employing the perturbation 
expansion with respect to

\begin{eqnarray}
H'=H(N)-H(N-1)={\partial H(N)\over
\partial N}+ ... .
\label{I0}
\end{eqnarray}
\noindent
Each additional derivative $\partial .../\partial N$
in the expansion (\ref{I0}) introduces a
factor $\sim 1/N$. Therefore, in the
limit $N\gg 1$, it is enough to consider the
first term only in (\ref{I0}). Correspondingly, 
calculations of the overlap are based on the $1/N$
expansion in Eq.(\ref{I0}) \cite{REM5}.
Below we follow the approach first 
introduced by Feynman and Vernon
in Ref.\cite{FEYNMAN} for describing weak interaction
with the bath. As pointed out in \cite{FEYNMAN},
it is sufficient to consider the effect of the
bath in the second order with respect to
this interaction, and the result must be 
considered as a first term of the expansion of 
the exponent. 
In our case the weakness is insured by the $1/N$
expansion. 
In other words, we look for the eigenstates ~$|m,N-1\rangle$~ 
of ~$H(N-1)$~ assuming that the eigenstates ~$|m,N\rangle$~
and the eigenergies ~$E_m(N)$ of ~$H(N)$~ are known. Accordingly,
in the lowest order with respect to
~$H'$~ (\ref{I0}) we find for (\ref{24})
 
\begin{eqnarray}
\chi_{m,m}=1- {1\over 2}\sum_{n\neq m}
{|(H')_{mn}|^2
\over \omega_{mn}^2 }, \quad \chi_{m,m'}= 
{(H')_{mm'}\over \omega_{mm'}} 
\label{I1}
\end{eqnarray}
\noindent
where the following notations

\begin{eqnarray}
(H')_{mn}= \widehat{\langle N, m|}H' \widehat{|n,N\rangle},
\quad \omega_{mn}=E_n(N)-E_m(N),
\label{J1}     
\end{eqnarray}
\noindent
have been introduced, and
the eigenstates $\widehat{|n,N\rangle}$
as well as the eigenenergies $ E_m(N)$ are
$exact$ with respect to the Hamiltonian
$H(N)$. 

As pointed out by Feynman and Vernon \cite{FEYNMAN},
if there is an extremely large number of 
bath degrees of freedom, the
exponentiation of (\ref{I1}) must be done after
the averaging. 
This is a consequence of the central-limit
theorem \cite{FEYNMAN}.
Thus, averaging ~$|\chi_{m,m}|^2$~ 
in Eq.(\ref{I1}) over the ensemble and then
exponentiating, we find the mean overlap 

\begin{eqnarray}
\overline{\chi}=\exp\left( -
\int_0^{\infty}
\,d\omega{J(\omega)\over
2\omega^2}\right)             
\label{I2}
\end{eqnarray}
\noindent
where the spectral weight $J(\omega)$
is defined as

\begin{eqnarray}
J(\omega)=\sum_{m, n}
p_m |(\delta H)_{mn}|^2
\delta(\omega - \omega_{mn}),    
\label{I3}
\end{eqnarray}
\noindent
and we have introduced the operator $\delta H$ whose
matrix elements are

\begin{eqnarray}
(\delta H)_{mn}
=(H')_{mn}- (H')_{mm}\,\delta_{mn}.  
\label{I5}
\end{eqnarray}
\noindent
We note that the accuracy of the result (\ref{I2})-
(\ref{I5}) is well controlled by the $1/N$ expansion
(\ref{I0}).

The OC occurs if the integral in (\ref{I2}) diverges
at the low frequency limit $\omega \to 0$. The borderline
for this divergence is the so called
ohmic dissipation \cite{BATH,LEG} characterized
by the limiting behavior ~$J(\omega)\sim
\omega^s$ ($s=1$) as $\omega \to 0$. For
$s>1$, the integral in Eq.(\ref{I2}) is finite,
which insures that no the OC occurs.
In the case when the bath spectral function
is ohmic (and subohmic) the OC
will occur, so that the condensate correlator will
irreversibly decay at large times.

Let us assume that no OC occurs, that is,
$\overline{\chi}\approx 1$ in Eq.(\ref{I2}).
Then, the correlator (\ref{3}) is represented by 
Eq.(\ref{10}). Its structure can be interpreted
in terms of the thermodynamic fluctuations
of the chemical potential 
~$\mu_m$. 
Employing the central limit theorem, we find 

\begin{eqnarray}
\displaystyle \rho^{(d)}_0 (t)=\overline{N}_0
{\rm e}^{
i\,\overline{\mu}t - (t/\tau_d)^2} + o(1)    
\label{E1}
\end{eqnarray}
\noindent
where the mean value of $\mu_m$ and the
dephasing time $\tau_d$ are given as 

\begin{eqnarray}
\overline{\mu}=\sum_mp_m(H')_{mm},\quad
\tau_d^{-2}={1\over 2}\sum_mp_m ((H')_{mm} - \overline{\mu})^2,
\label{E2}
\end{eqnarray}
\noindent
and the definition of ~$\mu_m$~ in Eq.(\ref{10}),
which can be written as 
~$\mu_m=\partial E_m(N)/\partial N + o(1/N)$~, 
as well as the identity $\partial E_m(N)/\partial N
=(H')_{mm}$, with
$H'$ defined by (\ref{I0}), have been employed. 

We note that the gaussian form (\ref{E1}) has its
limitation time $\tau'$, so that (\ref{E1})
is valid for $t \ll \tau'$. It is important to
realize that $\tau' \gg \tau_d$ in the macroscopic
system, so that when $t$ is larger than $\tau_d$ by
only few times the correlator (\ref{E1}) is
essentially zero. Therefore, the gaussian
approximation (\ref{E1}) is practically exact
(if no spontaneous revivals of the kind \cite{WALLS} are to
be expected to occur at times $\gg \tau_d$). 
We will discuss this $\tau'$ in detail later. 
 
Note that the overlap (\ref{I2}) can be interpreted
in a different manner. Indeed, transforming Eqs. (\ref{I3}),
(\ref{I2}) into the time representation
and introducing the spectral function $G(t)$ in the 
time domain as the inverse Fourier transform of
$J(\omega)$, one finds

\begin{eqnarray}
\overline{\chi}=\exp\left( - 
\int_0^{\infty} dt\, \int^t_0 \,dt' G(t')\right), \quad G(t)= 
\langle \delta H(t') \delta H(0)\rangle,
\label{I4}
\end{eqnarray}
\noindent
where the brackets denote the thermodynamic averaging
over the canonical ensemble with given $N$;
and ~$\delta H(t)=\exp[iH(N)t] \delta H\exp[-iH(N)t]$.
Thus, the OC is tightly connected with the 
long time behavior of the correlator $G(t)$. If $G(t)$
does not exhibit long-time correlations,
the conjecture of the presence of the OC implies
that $G(t)$ contains the white noise part, that is
$G(t)=\Gamma \delta (t) + G'(t)$, where $\Gamma >0$ is 
some effective decay constant, and $G'(t)$ stands for
the part which satisfies ~$\int_0^{\infty}\, dt
G'(t)=0$. The decay constant $\Gamma$ determines
the time scale $t_{OC}=1/\Gamma$ on which the OC
develops.
Evidently, the validity of the traditional treatment
of the bosonic system 
is limited 
by times ~$t\ll t_{OC}$~ (or by
the excitation frequencies $\omega \gg t_{OC}^{-1}$) 
\cite{RUK}
under the assumption that
the condensate correlator ~$\langle 
a_0^{\dagger}(t)a_0(0)\rangle =
\exp(-i\overline{\mu}t)$ \cite{STAT}. 

It is also possible to find 
the correlator (\ref{3}) as a $1/N$ cumulant expansion,
that is, $1/N$ expansion in the exponent.
In the main $1/N$ order, we find

\begin{eqnarray}
\displaystyle \rho_0(t)=\overline{N}_0\,
{\rm e}^{i\overline{\mu}t
-\left(t/\tau_d\right)^2}\,
\overline{\chi}(t);\quad
\overline{\chi}(t)=\exp\left( -
\int_0^t dt_1\, \int^{t_1}_0 \,dt' G(t')\right), 
\label{I40}
\end{eqnarray}
\noindent
where we have employed Eqs.(\ref{I1}),
(\ref{18}) in Eq.(\ref{3}), and, then,
performed the exponentiation up
to $o(1/N)$; the
quantities $\overline{\mu},\,\tau_d$
and $G(t)$ are defined by Eqs.(\ref{E2}),
(\ref{I4}). Here we have introduced 
$\overline{\chi}(t)$ which can be
interpreted as a time dependent overlap,
so that  
$\overline{\chi}=\overline{\chi}(t=\infty)$.

We emphasize that the gaussian
factor $\exp(-t^2/\tau_d^2)$ in the
form (\ref{I40}) {\it is not} due to the
short time approximation of the correlator
$G(t)$. As the expressions (\ref{E2})
and (\ref{I5}) indicate, the dephasing rate
$\tau_d^{-1}$ is determined by the
diagonal elements of the derivative ~$H'$~(\ref{I0})
of the $projected$ Hamiltonian ~$H(N)$~, while
the correlator $G(t)$ (\ref{I4}) is built
on essentially off-diagonal elements of $H'$.

The physical interpretation of (\ref{I40})
is the following. The condensate correlator
~$\rho_0(t)$~ is the ensemble mean
of the 
correlators ~$\langle N,m|a^{\dagger}_0(t)
a_0(0)|m, N\rangle$~ in the given
exact $N$-body eigenstates. 
These partial correlators contain the non-decaying
parts ~$\exp(i\mu_mt)$, characterized 
by the chemical potential $\mu_m$ 
 (defined in Eq.(\ref{10})). The correlators
~$\langle N,m|a^{\dagger}_0(t)
a_0(0)|m, N\rangle$~  
may also contain the decaying parts 
~$\exp (-\Gamma_m t)$,
characterized by some 
decay constants  ~$\Gamma_m >0$~. 
In general, there is no direct
relation between the parameters ~$\Gamma_m$~
and ~$\mu_m$, because ~$\mu_m$~ is determined
by the diagonal part of the operator ~$H'$~ (\ref{I0}),
while ~$\Gamma_m$~ are given by the off-diagonal 
elements of this operator.
The thermal averaging of the
partial correlators 
should not change the exponential decay much,
so that it can be characterized by a single 
exponent  
 ~$\sim \exp (-\Gamma t)$~ with some
mean decay constant $\Gamma$, which
can be related to $G(t)$ in (\ref{I4})
as ~$\Gamma \approx \int_0^{\infty} dt
G(t)$~ at times longer than some typical
correlation time of ~$G(t)$~.
It is important that the thermal averaging
produces the gaussian
time dependence in Eq.(\ref{I40})
due to the fluctuations of $\mu_m$
in the non-decaying factors  ~$\exp(i\mu_mt)$.
Thus, in general, one may expect the correlator
(\ref{I40}) to acquire the form

\begin{eqnarray}
\displaystyle \rho_0(t)=\overline{N}_0\,
{\rm e}^{i\overline{\mu}t
-\left(t/\tau_d\right)^2 - \Gamma t},
\label{I400}
\end{eqnarray}
\noindent
at times longer than some typical
correlation time of ~$G(t)$~. We note that
considering thermal (ensemble) fluctuations
of ~$\mu_m$~  makes sense, if 
the dephasing time $\tau_d$
is shorter than (or at least comparable to)
the OC time $t_{OC}=1/\Gamma$.  

As we will see later,
perturbative calculations yield
$t_{OC} \sim N$, while $\tau_d \sim \sqrt{N}$
in the case when no external bath is present.
In this case, the OC becomes irrelevant for all
practical purposes because ~$\rho_0(t)$~ of 
Eq.(\ref{I40}) can be well approximated by the
form (\ref{E1}) as long as $t\ll t_{OC}$,
and it will decay to almost zero 
at $\tau_d < t \ll t_{OC}$. 
Furthermore, we will discuss that, in the
isolated trap, the calculated  
~$t_{OC}$~ turns out to be longer 
than an inverse of the typical value
~$\Delta$~ of the matrix elements of the
interaction part of the many-body Hamiltonian. 
This implies that 
the
effect of the {\it level repulsion} should  modify 
significantly the correlator ~$G(t)$~ at
times ~$t\geq \Delta^{-1}$~. As will be
shown, the effect of the {\it level repulsion}
eliminates
completely the OC in the isolated condensate, 
so that the evolution is dominated
solely by the gaussian decay (\ref{E1}).

Thus, while the
necessary condition for irreversible
decay of the condensate correlator (\ref{3})
is the OC represented as 
$\overline{\chi}=\overline{\chi}(t=\infty)\to 0$,
a sufficient condition for the 
irreversible
decay is $ t_{OC}\leq \tau_d$, provided the
time scale on which the effect of the {\it level repulsion} 
may become significant is much longer than
~$t_{OC}$. This situation cannot be achieved in
the isolated trap, and some external bath 
characterized by its own corresponding spectral 
function is required in order to create
the OC. 
In the following discussion we will show that,
as long as an infinite interacting bath contacts a
finite condensate, the OC should always develop.

\section{Orthogonality catastrophe in the 
bosonic trap in the presence of an external bath}

First, we analyze the simplest model which exhibits
the OC. In this model we neglect
the normal excitations of the condensate, and take into
account excitations of some external bath, 
represented by
an infinite set of the non-interacting linear oscillators
\cite{FEYNMAN,BATH}. In this case, 
the {\it projected states} as well as the
spectral function $J(\omega)$ 
can be constructed explicitly. 
We will also analyze the role of the interaction
in the bath, and will demonstrate that 
the OC should occur in the interacting infinite bath
even though no OC was found 
without such an interaction. 
   
\subsection{Bath of non-interacting oscillators}

We neglect now excitations inside the trap
and assume a presence of some external bath
which, however, does not exchange particles with
the trap. For illustrative purposes,
the trap is represented by a single level
containing all the bosons. The bath is 
described by an infinite set of linear 
non-interacting oscillators \cite{FEYNMAN,BATH}.
This toy model is a simplified version of the model
suggested by Anglin \cite{ANG}.
Thus, we have 
the Hamiltonian  

\begin{eqnarray} 
H_0=(\epsilon_0+Q)a_0^{\dagger} a_0 +
\sum_a[{ p_a^2\over 2}   + {\omega_a^2 q_a^2\over 2}], 
\quad Q=\sum_a\left( g_aq_a + {g'_a\over 2}q_a^2\right),  
\label{I6} 
\end{eqnarray}
\noindent
where $\epsilon_0 $ stands for the energy of a
single level accommodating $N$ bosons; the
summation is performed over the non-interacting
oscillators, with $\omega_a$ and $g_a,\,\, g'_a$ 
being the frequency of the $a$th oscillator and 
the interaction constants, respectively.

The advantage of the model (\ref{I6}) is that
the eigenfunctions and the eigenenergies can be found
explicitly. Furthermore, the {\it projected states} can
be constructed explicitly as well.
The eigenstates and the eigenenergies of the Hamiltonian
(\ref{I6}) are

\begin{eqnarray} 
\displaystyle |m,N\rangle = 
|N\rangle \widehat{|m,N\rangle}, 
\quad \widehat{|m,N\rangle}=
\prod_a\Phi_{n_a}(q_a-g_a\Omega^{-2}_aN,\, \Omega_a),
\label{11} \\
E_m(N)=\sum_a\Omega_a(n_a+{1\over 2})
-{1\over 2}\left(\sum_a{g_a^2\over 
\Omega^2_a}\right)N^2,                  
\label{12}  
\end{eqnarray}
\noindent
where $|N\rangle$ stands for the Fock 
state of $N$ bosons occupying the
level
$\epsilon_0 $; $\Omega_a=\Omega_a(N)=
 \sqrt{\omega_a^2 + g'_aN}$, and we have explicitly
indicated the dependence of ~$\Phi_{n_a}$~ on
~$q_a$~ and ~$\Omega_a$.
The {\it projected states} 
$\widehat{|m,N\rangle}$, introduced in Sec.II,
are 
actually the bath states which
depend on $N$ as a parameter.
The state $\widehat{|m,N\rangle}$
is represented here in terms of the
oscillators eigenfunctions ~$ \Phi_{n_a}(q_a,\,
\Omega_a)$,
with $m$ carrying the meaning of the set
of all the integer quantum numbers $\{n_a\}$.
The $projected$ Hamiltonian $H(N)$ is, then,
obtained from (\ref{I6}) by simply
replacing $ a_0^{\dagger} a_0$ by $N$. Accordingly,
the derivative (\ref{I0}), which determines
the spectral weight and the chemical potential becomes

\begin{eqnarray} 
H'=H(N)-H(N-1)=\epsilon_0+ 
\sum_a\left( g_aq_a + {g'_a\over 2}q_a^2\right).
\label{I00}
\end{eqnarray}
\noindent
As discussed above in Sec.III, 
the diagonal elements of this derivative
determine the chemical potential in the given
eigenstate as ~$\mu_m=(H')_{mm} + o(1/N)$~ or

\begin{eqnarray}
\mu_{\{n_a\}}=\epsilon_0 +
\sum_a {g'_a \over 2\Omega_a   }
(n_a + {1\over 2})-N \sum_a{g_a^2\over
\Omega^2_a},
\label{AD2}
\end{eqnarray}
\noindent
where the label $m$ now carries the meaning
of the occupation numbers ~$\{n_a\}$~
of the oscillators. The averaging of
~$\exp (-i\mu_{\{n_a\}}t)\sim \prod_a
\exp (itg'_an_a/2\Omega_a)$~ over
$\{n_a\}$ produces the gaussian decay with the
dephasing rate ~$\tau_d^{-1}$~
following from Eq.(\ref{E2}) as

\begin{eqnarray}
\displaystyle
\tau_d^{-2}=\sum_a {g '^2_a \over \Omega^2_a}[
\overline{n}_a^2 +\overline{n}_a + {1\over 8}].
\label{160}
\end{eqnarray}
\noindent
Here $\overline{n}_a=(\exp(\beta \omega_a) -1)^{-1}$
stands for the mean thermal occupation
of the $a$-th oscillator. 
It is important to note that in the canonical
ensemble, the term ~$\sim g_a^2$~ in Eq.(\ref{AD2})
does not produce any dephasing. If, however,
~$N$~ fluctuates from realization to realization
(shot noise) and does not change during
the each realization, the dephasing
rate (\ref{160}) will acquire a contribution
given by the variance ~$\Delta N$~ of ~$N$~,
so that the result (\ref{160}) should be modified
as ~$\tau_d^{-2}\to \tau_d^{-2} + (\Delta N)^2
 (\sum_ag_a^2\Omega^{-2}_a)^2/2$~.

The
off-diagonal elements ~$(H')_{mn}$~ 
define the spectral weight ~$J(\omega)$~
which may lead to irreversible decay.
 Which one of these two effects --
either the reversible dephasing  
or the irreversible decay --  dominates
depends on the model parameters ~$g_a,\,\, g'_a$~
as well as on the density of states of the bath. 

Let us calculate the overlap (\ref{24})
directly  in order to see how the central
limit arguments work for an infinite
bath. This overlap is 
~$\widehat{\langle N-1,m}
\widehat{|m, N\rangle}=\prod_a\,\int\,dq_a\Phi^*_{n_a}
(q_a-g_a\Omega^{-2}_a(N-1)(N-1),\, \Omega_a(N-1))
\Phi_{n_a}( q_a-g_a\Omega^{-2}_a(N)N, \, \Omega_a(N))$~,
and it can be found explicitly. 
There is, however, no need to calculate it
exactly because the coefficients $g_a,\,\,g'_a$ must
be scaled with the number of the effective degrees
of freedom of the bath. Let us assume that the bath volume $V_b\to
\infty$ is
proportional to this number,
and the volume
occupied by $N$ bosons is fixed. Then, the summation
$\sum_a...$ introduces a factor $V_b$. In order to maintain
the energy an extensive quantity, we impose $g'_a\sim 1/V_b$,
and $g_a\sim 1/\sqrt{V_b}$. This implies that 
the overlap, which
can be represented as $\widehat{\langle N-1,m}
\widehat{|m, N\rangle}=\exp \{\sum_a\ln [\int\,dq_a\Phi^*_{n_a}
(q_a-g_a\Omega^{-2}_a(N-1)(N-1),\, \Omega_a(N-1))
\Phi_{n_a}( q_a-g_a\Omega^{-2}_a(N)N,\, \Omega_a(N))]\}$, can be
expanded in the inverse powers of $V_b$ in the exponent.
Accordingly, the only terms surviving the limit
$V_b\to \infty$ are linear in $g'_a$ and quadratic in $g_a$.
This is a reiteration
of the central-limit theorem \cite{FEYNMAN}. Thus, the mean
overlap (\ref{1000}) $\overline{\chi}$ acquires the form
(\ref{I2}),
with the
spectral weight 
 
\begin{eqnarray} 
J(\omega)=\sum_a{g_a^2\over \omega_a}
(\overline{n}_a +{1\over 2})\delta (\omega - \omega_a)
.                  
\label{14}  
\end{eqnarray}
\noindent
Of course, the
same expression (\ref{14}) 
follows from Eqs.(\ref{I3}), (\ref{I5})
where (\ref{I00}) must be employed in the limit $V_b
\to \infty$.

It is important that
the dephasing rate (\ref{160})
vanishes as ~$\tau_d^{-1}\sim 1/\sqrt{V_b} \to 0$~
in the limit ~$V_b\to \infty$~
(if no shot noise is taken into account).
The occurrence of the OC, which is
expressed as the condition
(\ref{26}), depends on 
the low frequency behavior of (\ref{14}).
If, e.g.,  $J(0)\neq 0$, the integral 
(\ref{I2}) will strongly diverge implying
 ~$ \overline{\chi} \to 0$. Thus, 
the OC is equivalent to the
irreversible dissipation for the considered model.   
Because, should the OC occur, its time $t_{OC}\approx 1/J(0)$
does not contain any positive power of the bath 
volume ~$V_b$~, so that ~$t_{OC} \ll \tau_d$~
in the limit considered.  
This corresponds to the irreversible loss of the phase 
memory on the time scale which
is finite in the thermodynamical limit. 
A concrete realization of this situation, requiring
an external bath, will be discussed elsewhere. 

If, however, the bath spectral function
does not lead to the OC (e.g., ~$J(\omega)\sim
\omega^2$~ as ~$\omega \to 0$~), the time
$t_{OC}$ is formally infinite, and the only
effect is the dephasing with ~$\tau_d \sim 
\sqrt{V_b}$. We note that the last situation
resembles most closely the situation in the
isolated trap, where the role of the bath
is played by the normal component confined
in the same trap.   

Concluding this section,
let us determine the validity of the gaussian
approximation (\ref{E1}) (or (\ref{I40})) with respect
to the given model in the case $V_b \to \infty$. 
Keeping in mind that 
$g'_a\sim 1/V_b$, as discussed above, 
the form (\ref{E1}) (or (\ref{I40}))  represents 
the first term in the exponent which is $\sim 1/V_b$.
Accordingly,
it is valid as long as the next term 
is much less than 1. 
The next term 
is $\sim \sum_a g_a '^3t^3\sim t^3/V_b^2$. It
becomes of the order of one, when $t\geq \tau'\sim
V_b^{2/3}$. Keeping in mind that $\tau_d \sim \sqrt{V_b}$
(see (\ref{160})),
we obtain that 
$\tau'\approx
\tau_d V_b^{1/6} \gg \tau_d$. Thus, as long as
the system is macroscopic, the approximation
(\ref{E1}) (or (\ref{I40}))
 is practically exact, as it was
discussed in Sec.III (see below Eq.(\ref{E2})).

\subsection{The OC induced by an external infinite bath 
of interacting degrees of freedom. The {\it strong OC}.}

Now we consider the role of the interaction
in the bath in producing the OC in the case
when no OC existed without interaction.
Referring to the previous model, let us assume
that (\ref{14}) 
 is such that no OC occurs.
What would happen, if the interaction between
$q_a$ is turned on? 
Such a question
is relevant for the confined
condensate. As it will be shown below, 
the intrinsic
bath of the non-interacting
normal excitations in the trap
does not produce the OC. Thus, it is important
to understand if the OC would emerge
when the interaction between
the normal excitations is taken into account.

Let us assume that the total Hamiltonian is ~$H=H_0+H_{int}$~,
where $H_0$ is the "free" part, and $H_{int}$
stands for some weak interaction. For example,
the free part can be given by $H_0$ (\ref{I6}), and
the interaction is
of the form $H_{int}= \sum_{abc}g_{abc} q_aq_bq_c $, with
$g_{abc}$ being the corresponding interaction constants. 
In fact, a concrete form of the interaction does not
matter much. We choose the third-power non-linearity
as an example which resembles most closely
the situation in the actual trap, where the most
important  interaction vertex contains 
three lines. 

The emergence of the OC in the presence of the
interaction can be viewed as an appearance
of the exponential decay in the correlator Eq.(\ref{I40})
leading to the suppression of the mean overlap
(\ref{I4}).
This happens if the interaction induces the decay
of $G(t)$ entering Eqs.(\ref{I40}),(\ref{I4}),
so that, e.g., $G(t)\approx 
\exp(-\gamma t)G(0)$ with some
$\gamma >0$. Then, the
correlator $\rho_0(t)$ (\ref{I40}) acquires the factor
$\exp(-\Gamma t)$ with $\Gamma=G(0)/\gamma$ at times
$t\gg 1/\gamma$. 
The gaussian factor $\exp(-t^2/\tau^2_d)$ in Eq.(\ref{I40}),
which is due to the ensemble fluctuations of the
chemical potential, 
is not sensitive to presence of the weak interaction
because the chemical potential may gain only small
corrections in the case $g_{abc}\to 0$. Thus, the
gaussian factor will not be significantly modified
by the interaction. Accordingly, the form
(\ref{I400}) should now describe the condensate correlator.

We take another look at the emergence of the OC.
In zeroth order with respect to $H_{int}$,
there is usually a high degree of degeneracy
between various multi-mode excitations.
This degeneracy is
represented by the condition ~$\omega^{(0)}_{mn}=
E^{(0)}_m(N)- E^{(0)}_n(N)=0$~ 
for some set $n\neq m$, where the energy levels
~$ E^{(0)}_n(N)$~ are found for $H_0$.
Here and below, the superscript ~$^{(0)}$~
denotes a quantity calculated in the
zeroth order with respect to ~$H_{int}$. 
Let us first consider the result of
the perturbation approach.
If the interaction 
$H_{int}$ is considered in the lowest order,
the matrix elements ~$(\delta H)^{(0)}_{mn}$~
(\ref{I5}) between 
some degenerate states may become finite
with some statistically significant weight. Then, if 
the energy levels are taken unrenormalized,
the summation close
to the degeneracy point will make
 $J(0)\neq 0$ in Eq.(\ref{I3}).
This yields
the OC because the integral (\ref{I2})
diverges as ~$\int_0^{\infty}\,d\omega{J(0)\over 
\omega^2}\to \infty$, implyng $\overline{\chi}\to 0$.
Accordingly, the time scale of the OC can be estimated
as ~$t_{OC}\approx J^{-1}(0)=\gamma /G(0)$.
       
It should be stressed, however, that
the role of the interaction $H_{int}$ is
two-fold. On the one hand, it opens transitions in
$\delta H$ (\ref{I5}) between the degenerate states,
and, on the other hand, it removes the degeneracy
by introducing the {\it level repulsion}. While the
first effect can be treated within the
perturbation expansion, the second one is essentially
non-perturbative. These two effects act in
the "opposite directions". The first one tends to
make $J(0)\neq 0$ even though originally it might be
that $J(0)=0$. The {\it level repulsion} removes
the degeneracy, and should result in $J(0)=0$.
Thus, whether the OC does or does not occur depends on 
how the {\it level repulsion} modifies the spectral weight in 
the limit ~$\omega \to 0$. 

The {\it level repulsion}
is presently extensively
discussed by the Random Matrix Theory (RMT) \cite{RMT}.
It has been shown \cite{RMT} that, due to the
{\it level repulsion}, 
the probability ~$ P(\omega)=
\langle \delta (\omega -\omega_{mn})\rangle$~
to find two eigenenergies $E_m$ and $E_n$
separated by the energy
"distance" $\omega$ exhibits the universal behavior
~$P(\omega)\sim \omega^s$~ for $\omega \to 0$, with 
$s=1,2,4$ depending on the general symmetry structure
of the theory \cite{RMT}. 
Recently, in Ref.\cite{CHAOS}
this feature has been explored to predict a dramatic
change in the dissipation in a chaotic system in 
the presence of a magnetic field.

We, however, note that it 
is possible to give simple physical arguments which
justify the irrelevance of the {\it level repulsion}
in the case when the external bath is infinite, while
the condensate is finite.
Indeed, the effect of the {\it level repulsion}
becomes significant when the energy difference
between two states is comparable (or less) with
a typical value of the interaction matrix elements
$(H_{int})^{(0)}_{mn}$ linking these states.
Then, some energy splitting 
$\Delta \sim |(H_{int})^{(0)}_{mn}|$ 
will occur. Otherwise, no {\it level repulsion}
should be taken into account.
In an infinite bath 
characterized by the
volume $V_b\to \infty$, the largest matrix elements
must be scaled as $(H_{int})^{(0)}_{mn}\sim V_b^{-\alpha }$,
where $\alpha >0$ is some power determined from
the requirement that the energy of the bath is
an extensive quantity. Thus, $\Delta\sim V_b^{-\alpha }\to 0$, 
and the {\it level repulsion} can be practically 
neglected because the physically relevant   
quantities, which describe the condensate,
cannot depend on ~$V_b\to \infty$~. Specifically, the time
$t_{OC}$, discussed in Sec.III, should be finite in this limit.
In this case,
no {\it level repulsion} should be taken into account, and 
the overlap ~$\overline{\chi}(t)$~ (\ref{I40}) 
becomes strongly suppressed 
as $\overline{\chi} (t)
\sim \exp(-t/t_{OC})\to 0$ as long as $t_{OC} < t \ll \Delta^{-1}$.
We will refer to such a situation as the {\it strong OC}.

In the case of the {\it strong OC},
the spectral function can be modeled as
$J(\omega)=J_0=const$
up to some high energy cut-off $\omega_c$.
The
constant $J_0$ can be restored from the normalization
condition ~$\int\, d\omega\, J(\omega)=\langle (\delta
H)^2
\rangle$~ obtained from Eq.(\ref{I3}). Thus,
$ J_0\approx \omega_c^{-1}\langle (\delta H)^2 \rangle$,
and, in accordance with the previous discussion,
the time scale $ t_{OC}\approx 1/J_0$, during which 
the OC actually occurs 
can be found as 

\begin{eqnarray}
t_{OC} = {\omega_c \over \langle
(\delta H)^2\rangle} \ll \Delta^{-1},
\label{J2}
\end{eqnarray}
\noindent
where the mean can be calculated in the zeroth
order with respect to ~$H_{int}$. 

Thus, a contact of a finite condensate 
with an infinite interacting bath 
should practically always result in  the {\it strong OC}.

\subsection{Absence of the {\it strong OC} in the 
condensate contacting
no external bath. The {\it Weak OC}.}

Now let us consider the case of the intrinsic bath,
that is, the normal component
confined together with the condensate (
for a moment, we neglect the possibility that the dephasing
effect may suppress the correlator
(\ref{I40}) much before the OC occurs).  
In this case, the {\it level repulsion} plays
an important role. 

In general, if the
operator ~$\delta H$~ entering Eq.(\ref{I3}) and defined by
Eqs.(\ref{I5}), (\ref{I0}) is not simply ~$\sim H_{int}$,
it is conceivable that ~$\delta H$~ links those degenerate
states (with statistical significance) which are not
mixed by ~$H_{int}$. In such case, these states
remain degenerate (up to the corresponding order).
Hence, the {\it level repulsion} would become
insignificant, and the OC may emerge. We note, however,
that such a case looks very artificial, when no
external bath is present, because
~$\delta H$~ is simply the derivative (\ref{I0})
of the $projected$ Hamiltonian. In the weakly interacting systems,
the $projected$ Hamiltonian can always be expanded in powers
of $N$. This automatically implies that ~$\delta H$~
is statistically dependent on $H(N)$. In other words, those
states which are linked by ~$\delta H$~ with statistical
significance are already mixed by the corresponding term
in $H(N)$, and their energies are thereby repelled
from each other.

As it will be seen in the next section,
the operator $\delta H$, which
determines the spectral weight (\ref{I3}), is
actually given as $N^{-1}H_{int}$, where
$H_{int}$ describes the interaction between 
the normal excitations. Thus, $H_{int}$
is responsible for turning on of the transitions
between the degenerate states as well as for
the {\it level repulsion}. The above arguments, 
which justified the irrelevance of the
{\it level repulsion}, when the external bath
is present, 
are not valid anymore for the intrinsic bath
because the typical
splitting $\Delta$ introduced above (see Sec.IVB)
is scaled now with $N$ instead of the infinite
volume of the external bath. 
 The value of $\Delta \sim |(H_{int})^{(0)}_{mn}|$
 stems from the
nature of interaction between the excitations.
The processes responsible for this interaction
are the pair scattering so that one member
of the pair 
either enters or leaves the condensate (for references
see \cite{REV} as well as the following discussion
in Sec.V).
Thus, $\Delta \sim 1/\sqrt{V}$ ( or $\sim 1/\sqrt{N}$
for fixed density). 

It is important to note that the OC time follows
from Eq.(\ref{J2}) as 
 $t_{OC} \sim N$, because now ~$\delta H= \delta H_{int}/N$~
is nothing else but the fluctuation of the interaction
energy per one particle. Accordingly, the system is now in the
limit $ t_{OC} \gg \Delta^{-1}$, which is the opposite
to the condition for the {\it strong OC}.
This implies, that the matrix elements $(\delta H)_{mn}$
as well as the excitation energies 
$\omega_{mn}$ in (\ref{I3}) experience strong renormalization
caused by the {\it level repulsion}. 

The above arguments imply that
the mean overlap (\ref{I2}) (or (\ref{I4}))
 in the isolated trap may be
sensitive only to the frequencies where the {\it level repulsion} 
takes place $\omega \leq \Delta$. This contrasts with 
the previous case of the external bath, in which the OC
occurs as the {\it strong OC}, that is,
far from the region where the {\it level repulsion} 
may become significant. Such a situation, when the OC occurs
due to the integration in
Eq.(\ref{I2}) over the region $\omega \leq \Delta$,
will be referred to as the {\it weak OC}. 

The actual
time scale $t_{OC}$ of such {\it weak OC} should always be
too long on any practical experimental time scale.
For example, if the {\it level repulsion}
modifies $J(\omega)$ from $J_0\neq 0$ to $J(\omega)\sim \omega$,
the divergence of the integral in (\ref{I2}) is logarithmic.
This immediately yields $t_{OC}\sim \exp (...N)\gg \Delta$, 
implying that $t_{OC}$ is practically infinity as long as
$N\gg 1$. Nevertheless, let us discuss a possibility of 
the {\it weak OC} in the isolated trap. 

As a matter of fact, $J(\omega)$ in (\ref{I2}) should 
approach zero not slower than ~$\sim \omega^2$ as $\omega \to 0$,
because the matrix elements 
~$(\delta H)_{mn}$~ in (\ref{I3}) are 
strongly renormalized by the
effect of the {\it level repulsion}. 
Indeed, if ~$|\tilde{n}\rangle$~
is a subspace of the degenerate states corresponding to
some energy ~$E^{(0)}_m(N)$~ (with respect to the
free Hamiltonian $H_0$), the new states and energies
can be found
by means of solving the corresponding secular problem
which includes this subspace only. Thus, 
the matrix ~$ (H_0)_{\tilde{n},\tilde{n}'}+
(H_{int})^{(0)}_{\tilde{n},\tilde{n}'}$~
must be diagonalized in the subspace of 
~$|\tilde{n}\rangle$~. The first part of this
matrix is ~$ E^{(0)}_m(N) \delta_{\tilde{n},\tilde{n}'}$~ 
due to the degeneracy. Thus, 
the new spectrum and the new states follow from
the diagonalization of 
~$(H_{int})^{(0)}_{\tilde{n},\tilde{n}'}$~ alone.
Accordingly, the renormalized 
~$(H_{int})_{\tilde{n},\tilde{n}'}$~ is 
$diagonal$ at the degeneracy. 
In fact, close to the degeneracy,
~$(H_0)_{mn}-E^{(0)}_m\delta_{mn}$~ plays
a role of the perturbation with respect to
~$(H_{int})^{(0)}_{mn}$.
This
implies that the renormalized
~$(\delta H)_{mn}\approx N^{-1}
(H_{int})_{mn} \to 0$~
as $\omega^{(0)}_{mn}\to 0$. 
The matrix elements must be smooth
in the parameters, therefore ~$\omega^{(0)}_{mn}\to 0$~ provides
a natural scale for the 
off-diagonal values of the renormalized matrix
elements ~$(\delta H)_{mn}$. 
Hence, the spectral weight (\ref{I3}), which is
 ~$\sim |(\delta H)_{mn}|^2$~
gains an 
extra factor $\omega^2$ in the limit $\omega \to 0$. 
This implies that the frequency 
integration in (\ref{I2}) over the region $\omega < \Delta $
does not yield any divergence. 
In  Appendix B, we will estimate the mean overlap
(\ref{I2})  
by employing the two-level approximation
for treating the {\it level repulsion}. The conclusion
is that this overlap $\overline{\chi}=1+o(1/\sqrt{N})$.

It is worth noting that the situation with
the irreversible decay of the condensate correlator
is very special with respect to the picture of the
decay of the normal excitations. Indeed, the decay time
of the long living normal excitations scales
with the typical size $L$ of the system, which
is $L\approx N^{1/3}$
in 3D.  It is much shorter than $\Delta^{-1}
\sim \sqrt{N}$. Thus, the normal excitations decay 
long before the {\it level repulsion} may
produce any effect. In contrast, the OC time ~$t_{OC}$~ 
calculated perturbatively with respect to
~$H_{int}$~ is $t_{OC} \sim N$, which is much longer
than $\Delta^{-1}
\sim \sqrt{N}$. Thus, the effect of the 
{\it level repulsion} is crucial for the long time dynamics
of the condensate.

The above analysis shows that neither the {\it strong OC}
nor the {\it weak OC}
should be anticipated in the isolated trap. 
The results discussed above are not
sensitive to particular details of the model.
Nevertheless,
below we will estimate explicitly 
the overlap ~$\overline{\chi}$~ within 
the hydrodynamic approximation.

To conclude this section, we emphasize that 
the question of the existence of the OC 
in a confined condensate is,
as a matter of fact, irrelevant 
as long as no external bath is present. 
Indeed, the earliest time when the OC may
become important is $t_{OC}$.
This time turns out to be $\sim N$ (see above).
On the other hand, the dephasing time
$\tau_d \sim \sqrt{N}$ (see Sec.VI). Therefore,
the correlator (\ref{I40}) will decay
long before the OC may become relevant.   
The mechanism of the decay of the
condensate correlator should be distinguished from 
that for the normal excitations, which
decay in 3D long before either the OC
or the dephasing may take place. 

\section{The overlap in the 
Hydrodynamic Approximation}

The notion of the OC \cite{OC} is well defined
in the thermodynamic limit $V\to \infty$
and $N\to \infty$. Strictly speaking, the 
$projected$ Hamiltonian $H(N)$ should be constructed
by employing the $N$-conserving approaches
\cite{GIR,GARD,CASTIN}. 
As was discussed
above, the occurrence of the OC is determined
by the low frequency (long time ) response of the 
excitation ensemble. 
Thus,
in order to construct
this response, it suffices to consider the
Hydrodynamic Approximation (HA) which 
is known as correctly reproducing the behavior of the
cloud at low frequencies and momenta
\cite{HYDRO}. The HA takes into
account correctly the conservation of the
total number of particles $N$ as well as
the vertex renormalization effects. A drawback
of the HA is the occurrence of
divergences at large frequencies. These are,
however, non-physical and can be taken
care of by introducing the upper cut-off
at some energy $\omega_c$ of the order of 
the chemical potential, and at some momentum
$k_c$ of the order of the inverse healing length.
On the contrary,
the divergences at low momenta and frequencies
are real and determine the peculiar physics
of the condensate \cite{HYDRO,REV}. Thus, in 
what follows we will employ the HA in order
to construct the $projected$ Hamiltonian
$H(N)$. 

\subsection{Dephasing due to non-interacting phonon gas}
We start with the 
classical Lagrangian  

\begin{eqnarray}
 L=\int d{\bf x} \Psi^*\left(
i\hbar { \partial \over \partial t}
+ {\hbar^2 \over 2m}\nabla^2 - U({\bf x}) +\mu -
{g\over 2}\Psi^*\Psi \right)
\Psi , 
\label{N1}  
\end{eqnarray}
\noindent
where the interaction term with $g=
{4\pi \hbar^2 a\over m} >0$ (
$a$ is the scattering length)
is taken in the contact form,  
with
$m$ being the atomic mass and $U({\bf x})$
standing for the trapping potential.
The chemical potential $\mu$ is introduced
to take into account the constraint

\begin{eqnarray}
\int d{\bf x} \Psi^*\Psi =N .
\label{N3}   
\end{eqnarray}
In the HA, the ansatz
 
\begin{eqnarray}
\Psi = \sqrt{n}{\rm e}^{i\theta}
\label{N4}   
\end{eqnarray}
\noindent
is employed in Eqs.(\ref{N3}), 
(\ref{N1}), where $n$, $\theta$ stand for the density and
the phase, respectively. This results
in the Hamiltonian

\begin{eqnarray}
H=\int d{\bf x} \left\{{\hbar^2 \over 2m}
n (\nabla \theta )^2 +(U-\mu)n +
{g\over 2}n^2\right\},  
\label{N5}
\end{eqnarray}
\noindent
where the terms containing gradients
of the density, which are not relevant
for the long wave dynamics, have been omitted.
We note that, if viewed together with
the constraint (\ref{N3}) in which $N$ is
just a c-number (not the operator), 
Eq.(\ref{N5}) represents the classical
version of the $projected$ Hamiltonian 
$H(N)$. Keeping this in mind, in what follows
we will use the notations $H$ and $H(N)$
interchangeably.

In order to construct the non-interacting
Hamiltonian $H_0$, the density is represented
as 

\begin{eqnarray}
n ({\bf x})= n_0({\bf x}) + n'({\bf x}), 
\label{N6}
\end{eqnarray}
\noindent
where $ n_0({\bf x})$ minimizes $H$ (\ref{N5}),
and saturates the constraint (\ref{N3}).
In the Thomas-Fermi approximation (see in \cite{REV}) 

\begin{eqnarray}
n_0({\bf x})= (\mu -U({\bf x}))/g, 
\label{N7}
\end{eqnarray}
\noindent
with the boundary condition 
$ n_0({\bf x})=0$ for $ U({\bf x})\geq \mu$.
Accordingly,
the constraint (\ref{N3}), which
becomes

\begin{eqnarray}
\int \,d{\bf x}n_0({\bf x})= N, 
\label{N77}
\end{eqnarray}
\noindent
yields the size of the condensate and the value
of $\mu$. 
Here we neglect the possibility of topological
excitations in the condensate. Thus, no winding
of the phase needs to be taken into account.

In Eq.(\ref{N6}), 
$n'({\bf x})$ denotes small perturbations 
(phonons) of 
the density which do not result in any change
of $N$. Thus, $\int d{\bf x}n'({\bf x})=0$.
Then, the Hamiltonian (\ref{N5}) takes the form
$ H=H_0+H_{int}$, where the quadratic part is 

\begin{eqnarray}
 H_0=\int d{\bf x} 
\left\{{\hbar^2 \over 2m}
n_0 (\nabla \theta )^2 +{g\over 2}
n'^2+{g\over 2}
n_0^2\right\},   
\label{N8} 
\end{eqnarray}
\noindent
provided (\ref{N7}) holds,
and the interaction is defined as

\begin{eqnarray}
H_{int}=\int d{\bf x} 
{\hbar^2 \over 2m}
n' (\nabla \theta )^2.   
\label{N9}
\end{eqnarray}
\noindent

Subsequent quantization of the
Lagrangian (\ref{N1}) shows that
the density $n' ({\bf x})$ and the phase
$\theta ({\bf x})$ are the conjugate variables.
The phonon contributions
can be described in terms of the 
Fourier harmonics  $\theta_{\bf k}$ and
$n'_{\bf k} $ of $\theta ({\bf x})$
and $n'({\bf x})$, respectively
(~$\theta ({\bf x})=\sum_{\bf k}{1\over \sqrt{V}}
\exp( i{\bf kx})\theta_{\bf k}$~, ~$n' ({\bf x})=
\sum_{\bf k}{1\over \sqrt{V}}
\exp( i{\bf kx})n'_{\bf k}$), satisfying
the commutation relation

\begin{eqnarray}
[\theta_{\bf k}, n'_{{\bf k}'}]=-i
\delta_{{\bf k},{\bf k}'}.  
\label{N11}
\end{eqnarray}
\noindent
Thus, one can express $ n'_{\bf k}=i 
{\partial \over \partial \theta_{\bf k}}$. The 
corresponding Hamiltonian is actually the
$projected$ Hamiltonian $H(N)$, in which 
$N$ plays the role of a parameter through
$n_0$ given by Eqs.(\ref{N7}), (\ref{N77}). 
The eigenstates of $H(N)$ are the
{\it projected states} $\widehat{|m,N\rangle}$, which
can be expressed as some functionals
$\Phi_m(\{\theta_{\bf k}\}, n_0)$. 

For sake of simplicity, we will consider
a finite cubic box with the side $L$, and 
will set the trapping potential to 0, 
with, e.g., periodic boundary conditions. 
Then, $n_0({\bf x})$ becomes just a constant
$n_0=N/V,\, V=L^3$, and
$H_0$ in Eq.(\ref{N8}) can be represented
as 

\begin{eqnarray}
\displaystyle
H_0=\sum_{{\bf k}}\left(-{g\over 2}{
 \partial^2 \over
\partial \theta_{{\bf k}}\partial \theta_{-{\bf k}}}\,\, 
+{\hbar^2  k^2n_0\over 2m}
\theta_{\bf k}\theta_{-\bf k} \right)\,
+{gN^2\over 2V},    
\label{N12}
\end{eqnarray}
\noindent
where 
$\theta_{\bf k}=\theta^*_{-\bf k}$. 
Taking into account Eqs.(\ref{N8}),
(\ref{N9}), $H'$ in (\ref{I0}) is 
given as 

\begin{eqnarray}
\displaystyle
H'=\partial H(N)/\partial N=
\sum_{{\bf k}}{\hbar^2  k^2\over 2Vm}
\theta_{\bf k}\theta_{-\bf k} \, +{gN\over V}\,\,\,.
\label{N120}
\end{eqnarray}
\noindent
Here the first term represents the global kinetic energy
density (notice the factor $V$ in the
denominator) of the phonon gas. This term 
contributes to the dephasing rate ~$\tau_d^{-1}$~ (\ref{E2})
at finite temperatures. The second term, which is the mean
interaction energy per one boson, does not contribute
to the dephasing in the canonical ensemble.
However, in case the evolution of the
correlator (\ref{1}) is considered for the case when the initial
state is a mixture of states with different $N$ (or
the averaging is performed over different realizations of $N$), 
the last term in Eq.(\ref{N120}) may contribute to the dephasing
rate (\ref{E2}). We will discuss this later with respect to
the effect of the shot noise inevitably present in the destructive
measurements \cite{JILA}.

Let us neglect for a while the last terms
in Eqs. (\ref{N12}),(\ref{N120}). Then, comparison of 
Eq.(\ref{N120}) with Eq.(\ref{I00})
indicates that the condensate
and the normal component can be viewed
as the case $g_a=0,\,\omega_a=0,\, g'_a\neq 0$ of the
toy model considered in Sec.IVA. Specifically,
in the absence of the interaction between
phonons, the $projected$ Hamiltonian (\ref{N12}) obtained
above in the HA is practically the $projected$ Hamiltonian 
of the form (\ref{I6}), with $g_a=0$ and
~$g'_a={\hbar^2  k^2\over 2Vm}$,
~$\Omega_a = \sqrt{n_0g/m}\,\,k$. The
spectral weight (\ref{I3}) then follows as ~$J(\omega)\approx
\sum (g '^2_a/\Omega^2_a)\overline{n}_a^2\delta (\omega
-\Omega_a)\sim \omega^2 T^2/ N $~
at large $T$, where Eqs.(\ref{I5}), (\ref{N120}),
(\ref{N12}) have been employed. 
Thus, no OC occurs due to the ideal
phonon gas because ~$J(\omega) \sim \omega^2/N$,
so that the integral in the overlap (\ref{I2}) does not diverge
for $\omega \to 0$. Furthermore, due to $J(\omega) \sim 1/N$,
the overlap ~$\overline{\chi}=1+o(1/N)$. Consequently,
the correlator (\ref{3}) is given by the form (\ref{E1}), where
the dephasing rate
given by Eq.(\ref{160}) turns out to be 
 $\tau^{-1}_d\sim T/\sqrt{N}$ (in Sec.VI, we will
present accurate calculations of $\tau_d^{-1})$. 
The following discussion will be devoted to
proving that the above simple picture is not 
altered practically
by the interaction between phonons.

\subsection{ Interacting phonon gas}

Now let us consider the role of the interaction between 
the phonons.
It is convenient to remove the dependence
on $n_0$ from ~$H_0$~ (\ref{N12}) and transfer it to the
interaction part (\ref{N9}). We employ the scaling 
transformation $\theta_{\bf k}=\lambda^{-1}\theta'_{\bf k}$,
with $\lambda =(n_0/n_r)^{1/4}$ where $n_r$ stands for some
reference density which will be kept constant with respect
to changing $N$. As a result
of this transformation, ~$H_0$~ (\ref{N12}) changes to 
$\lambda^2H'_0$
(we neglect the last term in Eq.(\ref{N12})),
where $H'_0$ has the form (\ref{N12}) in which $\theta_{{\bf k}}$
is replaced by $\theta'_{{\bf k}}$ and $n_0$ is replaced 
by $n_r$. This scaling transformation will change the interaction
part (\ref{N9}) as $H_{int}\to \lambda^{-1}H_{int}$. Thus, the total
Hamiltonian can now be represented in the new variables as

\begin{eqnarray}
\displaystyle
H&=&\left({n_0\over n_r}\right)^{1/2}(H'_0+H'_{int}),
\label{N13}
\\
\displaystyle H'_0&=&  
=\sum_{{\bf k}}\left(-{g\over 2}
{ \partial^2 \over
\partial \theta'_{{\bf k}}\partial \theta'_{-{\bf k}}}\,\, 
+{\hbar^2  k^2n_r\over 2m}
\theta'_{\bf k}\theta'_{-\bf k} \right)
\label{N14}
\\
\displaystyle H'_{int}&=&\left({n_0\over n_r}
\right)^{-3/4}{i\hbar^2\over 2m\sqrt{V}} \sum_{{\bf k},{\bf q}}
{\bf kq} \theta'_{{\bf q}}{\partial \over
\partial \theta'_{-({\bf k + q})}}\left(\theta'_{\bf k} ...\right).  
\label{N15}
\end{eqnarray}
\noindent
In the representation (\ref{N13})-(\ref{N15}), the matrix
elements of the operator
$\delta H$, which determines
the spectral weight in Eq.(\ref{I3})
and which is given by Eqs.(\ref{I5}), (\ref{J1}), (\ref{I0}), 
become

\begin{eqnarray}
(\delta H)_{mn}=-{3\over 4N}\left(( H'_{int})_{mn}-
\delta_{mn}( H'_{int})_{mm}\right).   
\label{N16}
\end{eqnarray}
\noindent
It is important to notice the factor
$1/N$ in this equation.

The meaning of the
scaling transformation introduced above
must be discussed. 
The overlap (\ref{24})
is calculated between two families of the eigenstates.
Specifically, the change $N\to N - 1$ produces
the change of $n_0=N/V$ as $n_0\to n_0 - 1/V$.
Thus, the overlap is to be found between the states
~$\Phi_m(\{\theta_{\bf k}\}, n_0-V^{-1})$~ and
~$\Phi_m(\{\theta_{\bf k}\}, n_0)$.
We represent the overlap (\ref{24}) explicitly
as an integral 

\begin{eqnarray}
\chi_{m,m} = \int D\theta_{\bf k}
\Phi^*_m(\{\theta_{\bf k}\}, n_0-V^{-1})
\Phi_m(\{\theta_{\bf k}\}, n_0)  
\label{N17}
\end{eqnarray}
\noindent
over all the harmonics
$\{\theta_{\bf k}\}$.
The eigenstates ~$\Phi_m(\{\theta_{\bf k}\}, n_0)$~  
can be expressed in terms of the eigenstates ~$\Phi'_m
(\{\theta'_{\bf k}\}, n_0)$~ of the new Hamiltonian ~$
H'_0+H'_{int}$~ (\ref{N13})-(\ref{N15}) (where 
~$\theta_{\bf k}=\lambda^{-1}\theta'_{\bf k}$~ and
~$\lambda =(n_0/n_r)^{1/4}$).
 Specifically,
~$\Phi_m(\{\theta_{\bf k}\}, n_0)
=\lambda^{M/2}\Phi_m'
(\lambda \{\theta_{\bf k}\}, n_0)$, where 
$M$ stands for the dimension
of  ~$ D\theta_{\bf k}=\prod_{\bf k}d\theta_{\bf k}$~. 
Then, Eq.(\ref{N17}) takes the form

\begin{eqnarray}
\chi_{m,m} =\left(1+{\delta \lambda \over \lambda}
\right)^{M/2}
\int D\theta_{\bf k}\,\,
\Phi'^*_m((1+{\delta \lambda \over \lambda})
\{\theta_{\bf k}\}, n_0-V^{-1})
\Phi'_m(\{\theta_{\bf k}\}, n_0), 
\label{N19}
\end{eqnarray}
\noindent
where $\delta \lambda $ is the change
of $\lambda $ as $n_0$ changes by $1/V$.
In fact, it is enough to consider only
the term lowest in $1/N$, so that 
~$\delta \lambda/\lambda = -(4N)^{-1}$.
Expanding (\ref{N19}) in $1/N$ and 
exponentiating
the result \cite{FEYNMAN} within the 
accuracy $N^{-2}$, we find
the logarithm of the mean overlap as

\begin{eqnarray}
\ln \overline{ \chi}
 =- 
{M\over 4(4N)^2} - B_0(T) -B_1(T)- B_2(T)
\label{N20}
\end{eqnarray}
\noindent
where the terms linear in $1/N$ 
cancel automatically, and the terms 
$B_0(T)$, $B_1(T)$, $ B_2(T)$
represent the following contributions 

\begin{eqnarray}
B_0(T)={1\over 2}\left({1 \over 4N}\right)^2
\sum_m p_m \int D\theta_{\bf k}\,\,
\sum_{\bf k,k'}\Phi'^*_m(
\{\theta_{\bf k}\}, n_0) \theta_{\bf k}
\theta_{\bf k'}{\partial^2 \over \partial \theta_{\bf k}
\partial \theta_{\bf k'}}
\Phi_m'( \{\theta_{\bf k}\}, n_0), 
\label{N190}
\end{eqnarray}
\noindent
 
\begin{eqnarray}
B_1(T)=4n_0\left({1 \over 4N}\right)^2
\sum_m p_m \int D\theta_{\bf k}\,\,
\sum_{\bf k}\Phi'^*_m(
\{\theta_{\bf k}\}, n_0) \theta_{\bf k}
{\partial^2 \over \partial n_0 
\partial \theta_{\bf k}}
\Phi_m'( \{\theta_{\bf k}\}, n_0), 
\label{N191}
\end{eqnarray}
\noindent

\begin{eqnarray}
B_2(T)=- 8n_0^2
\left({1 \over 4N}\right)^2
\sum_m p_m \int D\theta_{\bf k}\,\,
\Phi'^*_m(
\{\theta_{\bf k}\}, n_0) 
{\partial^2 \over \partial n^2_0} 
\Phi_m'( \{\theta_{\bf k}\}, n_0). 
\label{N192}
\end{eqnarray}
\noindent
The first term in (\ref{N20}) contains the formally
divergent dimension $M$. This can be
represented as ~$M=\sum_{\bf k}\sim V\int d^3k \to \infty$.
The divergence of the integral ~$\int d^3k$~ is a
general high momenta divergence of the HA, and therefore
it is
non-physical. The integral
must be cut off from above at the inverse
healing length $k_c$. Thus, $M\sim N$, and
the first term in (\ref{N20}) behaves
as $\sim 1/N \to 0$, and therefore it
can
be eliminated. 

The term $B_0(T)$ represents
a contribution which is finite in the absence
of the interaction (\ref{N15}). Thus, it originates from
the free phonons as discussed above in
Sec.VA. Accordingly, elementary gaussian 
calculations with the free Hamiltonian
(\ref{N14}) yield   

\begin{eqnarray}
B_0(T)=
{1\over2}\left({1 \over 4N}\right)^2
\sum_{\bf k}{\exp({\hbar \omega_{\bf k}\over T})
\over (\exp({\hbar \omega_{\bf k}\over T})-1)^2}, 
\label{N30}
\end{eqnarray}
\noindent
with $\hbar\omega_{\bf k}\sim k $ being the spectrum of the normal
excitations of the free Hamiltonian (\ref{N14}). 
The term (\ref{N30}) is $\sim 1/N$, and it converges
at small $k$ in the 3D case. The
interaction may slightly change the "free"
form (\ref{N30}) without changing the main 
scaling dependence on $N$. Thus, the term (\ref{N190})
can be eliminated from the consideration as well. 

The term (\ref{N191}) describes a contribution
which is non-singular at the degeneracy points
$\omega_{mn}$ (see the discussion in Sec.IV).
The proof of this is presented in Appendix
A. Thus, 
the term (\ref{N191})
contributes as $\sim 1/N$
in the overlap (\ref{N20}). Consequently,
we eliminate $B_1(T)$ as well.

Finally, the term (\ref{N192}) is the one which
is formally singular at the degeneracy points. Thus, 
one can represent $\overline{\chi}
=\exp (-B_2(T))$. It is possible to
recognize (see the Appendix A)
this form as Eq.(\ref{I2})
with the spectral weight

\begin{eqnarray}
J(\omega)=
\left( {3\over 4N}\right)^2
\sum_{m, n\neq m}
p_m |( H'_{int})_{mn}|^2
\delta(\omega - \omega_{mn}),    
\label{N22}
\end{eqnarray}
\noindent
where $H'_{int}$ is given by Eq.(\ref{N15}).
It is important to note the factor ~$1/N^2$~
in Eq.(\ref{N22}).

Above we have justified that the $projected$
Hamiltonian $H(N)$ as well as its 
space of states can be expressed in such a way
that the dependence on $N$ is included in
the interaction part $H'_{int}= H'_{int}(N)$
(\ref{N15}) only. This implies that,
should the OC occur, solely the interaction 
would be responsible for this.
It is worth noting that in (\ref{N22}), the matrix
elements are taken between $exact$ eigenstates
of the "primed" Hamiltonian $H'_0+H'_{int}$
(\ref{N13}).

Eq.(\ref{N22}) contains
usual HA high momenta divergences.
Such divergences, which are non-physical,
can be eliminated
by the following procedure.
Let us replace $\overline{\chi}\to
\overline{ \chi}/\chi_{0,0}$, where $\chi_{0,0}$
stands for the overlap of the $projected$
ground states $\widehat{|0,N-1 \rangle}$
and $\widehat{|0,N \rangle}$. 
It has been shown that such
an overlap is essentially 1 (see the discussion
below Eq.(\ref{25})). Thus, the
renormalized overlap (\ref{I2}) becomes

\begin{eqnarray}
\overline{ \chi}
 =
\exp\left( -
\int_0^{\infty}
\,d\omega{J(\omega,T) - J(\omega,0) \over
2\omega^2}\right) ,
\label{N21}
\end{eqnarray}
\noindent
where the temperature dependence is shown
explicitly  in $J(\omega)= J(\omega, T)$.
Eqs.(\ref{N21}), (\ref{N22}) still have
divergences at high momenta, which, however,
vanish at $T=0$. These divergences must be
cut off from above at the momentum 
~$k_c\approx \sqrt{an_0}$~
\cite{REV,HYDRO}.

Let us first make naive estimates of 
$t_{OC}$ from Eq. (\ref{J2}). That is,
we ignore the {\it level repulsion} effect,
and consider the form (\ref{J2}) in the lowest
order with respect to the perturbation
theory, where ~$\delta H$~ is taken from
Eq.(\ref{N16}).
Thus, we obtain

\begin{eqnarray}
t_{OC}\approx { N^2 \omega_c\over \sum_{m, n\neq m}p_m
|(H'_{int})^{(0)}_{mn}|^2 }
\label{N210}
\end{eqnarray}
\noindent
where Eq.(\ref{N15}) should be employed.
The denominator of Eq.(\ref{N210}) is the
mean square total interaction energy 
between the phonons. Thus, it is scaled
as $\sim N$ as any extensive
quantity, and Eq.(\ref{N210})
yields $t_{OC}\sim N$. On the other hand,
the typical matrix element ~$\Delta \sim (H'_{int})^{(0)}_{mn}$~
(see discussion in Sec.IVC), where ~$H'_{int}$~ is given by
Eq.(\ref{N15}), is scaled as ~$\sim 1/\sqrt{N}$. Thus,
 ~$t_{OC} \gg \Delta^{-1}$, and the 
{\it strong OC} cannot occur
(see discussion in Sec.IVB).
We note that 
the dephasing time $\tau_d \sim
\sqrt{N}$ (see below Eq.(\ref{N120})).
Thus, regardless of whether
the {\it weak OC}
does or does not formally occur, the dephasing
effect dominates the 
evolution (\ref{I40}) (because $\tau_d \ll t_{OC}$ in the
limit $N\gg 1$). This implies that the question
whether the OC actually occurs 
becomes irrelevant because the correlator
(\ref{I40}) decays at times $\geq \tau_d$
which are much shorter than the time ~$t_{OC}$~ 
(\ref{N210}) of the
possible OC. Nevertheless, in Appendix B,
we have discussed
the role of the
{\it level repulsion}, and show that no the 
OC occurs. Furthermore, in Appendix B, we will
evaluate the mean overlap $\overline{\chi}$, and 
it will be shown that the only effect caused by the
interaction between the phonons is that 
~$\overline{\chi}=1+o(1/\sqrt{N})$, while in the
non-interacting phonon gas ~$\overline{\chi}=1+o(1/N)$
as it was concluded in Sec.VA.   

Thus, in the case of no extrinsic bath,
the decoherence
of the condensate correlator is essentially
given by Eq.(\ref{E1}). Below we will estimate the
dephasing time for the box.

\section{The dephasing time of a confined condensate}

Here we will calculate the dephasing
rate $\tau_d^{-1}$ of the condensate time-correlator
as described by Eq.(\ref{E1}).
The chemical potential $\mu_m=(H')_{mm}$~, where
~$H'$~ is given by Eq.(\ref{N120}), 
consists of two parts $\mu^{(0)}$
and $\mu'_m$. The term $\mu^{(0)}=gN/V$ 
results from the last term in Eq.(\ref{N8}),
(\ref{N12}), (\ref{N120}),
and it is not associated with excitations. In the canonical
ensemble, it does not produce any contribution
to ~$\tau_d^{-1}$~ in (\ref{E2}). We note, however, that 
in the destructive measurements \cite{JILA}, ~$N$~
will necessarily fluctuate
from the realization to the realization
producing the shot noise. Therefore, 
the term $\mu^{(0)}$ will contribute into
the dephasing rate ~$\tau_d^{-1}$~ (\ref{E2}).
To estimate this effect, we assume that
the shot 
fluctuations are characterized by some known
variance $\Delta N$. This 
yields the dephasing rate due to the shot noise as

 \begin{eqnarray}
(\tau^{(0)}_d)^{-1}={g\Delta N\over \sqrt{2}V}.
\label{E22}
\end{eqnarray}
\noindent
It is important to emphasize that the shot
noise, which is due to the uncertainty of 
$N$, does not disrupt the phase
coherence in the each realization. It should not
be identified with the situation when $N$ is not conserved
due to the unwanted escape (or deposition) of 
particles from (to) the trap
during the evolution. The latter effect is a source of
the white
noise which erases any memory, and which leads to
the exponential decay instead of the gaussian dephasing
(\ref{E1}) \cite{SOL}.   

The term $\mu'_m$ is due to
the normal component. In order
to obtain it, the spectrum of the
elementary 
excitations $\epsilon_{\bf k}(N)$ must be found.
According to the
preceding analysis, 
the processes of
scattering (elastic and inelastic)
of the excitations do not contribute
to the correlator (\ref{3}) in the leading
order $N\gg 1$. Thus, in order
to find $\tau_d$ in (\ref{E2}), the 
interaction between the excitations can be
essentially ignored. 
This will give 
the energy $E_m(N)$ as 
~$E_m(N)=\sum_{\bf k}\epsilon_{\bf k}(N)n_{\bf k}$,
where $ n_{\bf k}$ are the corresponding
population factors of the elementary
excitations. 

Strictly speaking,
in order to find the effect ($\tau_d^{-1}\neq 0$)
for $N$ finite, 
the $N$-conserving approaches \cite{GIR,GARD}
should be employed. Above we have employed 
the HA, which is the long wave limit of
these approaches. The HA has been
sufficient for estimating the mean
overlap in order to justify the absence of
the OC. However, here we are interested in 
accurate calculations of the dephasing
time $\tau_d$ from Eq.(\ref{E2}). These
calculations contain the integration 
over the momenta which are divergent on the
high end, and which therefore cannot
be accurately eliminated within the HA
approach. Thus, the methods \cite{GIR,GARD}
is the only known alternative. However,
we note that in the limit $N\gg 1$,
the differences vanish between the approaches
\cite{GIR,GARD} and the non-conserving
$N$ Bogolubov theory \cite{BOG}.
This implies that, if we employ the
approaches \cite{GIR,GARD} in evaluating
the elementary excitation spectrum, we
will gain corrections of the order
$1/N$ compared with the Bogolubov
method. Keeping in mind that the
result in (\ref{E2}) is itself $\sim 1/N$,
the corrections
to the spectrum will produce terms 
$\sim 1/N^2$,
which should be neglected. 

Thus, in what 
follows we will employ the Bogolubov
method \cite{BOG}, and take 
$\epsilon_{\bf k}$ as the standard Bogolubov spectrum
(see \cite{BOG,REV}).
We note that the form (\ref{10}) indicates that,
 while changing $N$ by 1, the number of excitations
$n_{\bf k}$ must be maintained unchanged. Thus, neglecting terms
$o(1/N^2)$ in Eq.(\ref{E2}), we obtain

\begin{eqnarray}
\displaystyle \tau_d^{-2}={1\over 2}\langle \left(\sum_{\bf k}
{\partial \epsilon_k 
\over \partial N}n_{\bf k} - \langle \sum_{\bf k}
{\partial \epsilon_{\bf k} 
\over \partial N}n_{\bf k}\rangle \right)^2\rangle.
\label{D1}
\end{eqnarray}
\noindent
It is important to emphasize that the thermodynamic
fluctuations, over which the averaging $\langle ... \rangle$
is to be performed in Eq.(\ref{D1}), affect not only 
the term containing the population factors $n_{\bf k}$
explicitly. Indeed, as has been
discussed by Giorgini et. al \cite{FLUC}, the condensate
fraction $N_0$ exhibits anomalous fluctuations $
\langle \delta N_0^2 \rangle$ . 
Consequently, these fluctuations must affect 
the spectrum of the normal excitations $\epsilon_{\bf k}$
depending on $N_0$. Below we will see that 
the contribution due to these fluctuations can be dominant
in
Eq.(\ref{D1}) for large $N$ and not very low $T$. 
We represent $\epsilon_{\bf k}=\epsilon_{\bf k}(N_0)
=\epsilon_{\bf k}(\overline{N}_0 +\delta N_0)$
as an expansion in a smallness quantity $\sqrt{
\langle \delta N_0^2 \rangle}
/{\overline N}_0 \ll 1$ around the mean value
${\overline N}_0$ of $N_0$. This gives 

\begin{eqnarray}
\displaystyle {\partial \epsilon_k(N_0)
\over \partial N}= {\partial  
\epsilon_k(\overline{N}_0)
\over \partial N}
+{\partial 
\epsilon_k(\overline{N}_0)
\over \partial \overline{ N}_0}
{\partial \delta N_0 \over \partial N}
+ {\partial^2 
\epsilon_k(\overline{N}_0)
\over \partial  N \partial \overline{N}_0} \delta N_0
+o(\delta N_0^2/\overline{N}^2_0). 
\label{D2}
\end{eqnarray}
\noindent
The fluctuation ~$\delta N_0$~ is understood here
as happening at fixed $N$, thus 
one can employ ~$\partial\delta N_0/\partial N=
\delta \partial N_0/\partial N$~.    
We note that $\partial \overline {N}_0/\partial N$ can be taken
as $\partial \overline {N}_0/\partial N=1$, because
the corrections due to the interaction contribute to the
higher order in the gas parameter $\xi$. 
Both this expansion and $n_{\bf k}=\overline{n}_k
+\delta n_{\bf k}$, where $\overline{n}_k$
stands for the mean of $n_{\bf k}$, should be substituted 
into Eq.(\ref{D1}). In calculating the average, only the lowest
terms in the fluctuations ($\langle \delta n^2\rangle$,
$\langle \delta N_0^2\rangle $, etc.) must be considered.
For large $N$, the differences between
canonical and grand canonical ensembles practically disapper
with respect to the normal component.
Thus, we
employ $\langle ( n_{\bf k} - \langle n_{\bf k} \rangle )^2
\rangle = \overline{n}_{\bf k} ( \overline{n}_{\bf k} +1)$,
where 

\begin{eqnarray}
\overline{n}_k={1\over 
\exp({\epsilon_k\over T})-1}.  
\label{F6}
\end{eqnarray}
\noindent

We neglect the terms of the order higher 
than $1/N$ in Eq.(\ref{D1}) (e.g.,
$\langle \delta N_0 n_{\bf k}
\rangle$).
Close examination shows that each derivative
of $\epsilon_k$ in Eq.(\ref{D2}) introduces
an additional power of the gas parameter
after the momentum summation in Eq.(\ref{D1}).
Thus, we retain only the first two terms
in the expansion (\ref{D2}). 
To proceed, we employ the approach
\cite{FLUC} in calculating the anomalous fluctuations.
Thus, we choose the Bogolubov
representation 
\cite{BOG,REV} of the normal component as

\begin{eqnarray}
\psi'&=&{1\over \sqrt{V}}
\sum_{\bf k}(u_{\bf k}\alpha_{\bf k} +
v_{\bf k}\alpha^{\dagger}_{-\bf k}){\rm e}^{i{\bf kx}},
\label{D3} \\
u^2_k+v^2_k&=&{(\epsilon_k^2+g^2n_c^2)^{1/2}\over
2\epsilon_k },
\label{D33}
\\ 
u_kv_k&=&-{gn_c\over 2\epsilon_k},
\label{D333}
\end{eqnarray}
\noindent
where $\alpha_{\bf k}$ ($\alpha^{\dagger}_{\bf k}$)
destroys (creates) a quasiparticle with the 
momentum ${\bf k}$; $n_c=N_0/V$ stands for the condensate density,
and
the Bogolubov spectrum is

\begin{eqnarray}
\epsilon_k=\left[\left({\hbar^2k^2 \over 2m} + gn_c\right)^2
-g^2n_c^2\right]^{1/2}.
\label{D4}
\end{eqnarray}
\noindent

Fluctuations of the condensate population
$\delta N_0$ are related to the fluctuations
of the normal component by means
of the relation  \cite{FLUC} 

\begin{eqnarray}
 N_0=N -\int d{\bf x} 
\psi '^{\dagger} \psi' \, . 
\label{D5}
\end{eqnarray}
\noindent
Accordingly, employing Eqs.(\ref{D3})-(\ref{D333}) and $n_{\bf k}=
\alpha^{\dagger}_{\bf k}\alpha_{\bf k}$ in Eq.(\ref{D1}),
we obtain

\begin{eqnarray}
\displaystyle \tau_d^{-2}&=&\tau_1^{-2}+\tau_2^{-2}
, 
\label{F1}
\end{eqnarray}
\noindent
where the notations are

\begin{eqnarray}
\tau_1^{-2}&=& {1\over 2}
 \sum_{\bf k}
\left({\partial \epsilon_k(\overline{N}_0)
\over \partial \overline{N}_0}\right)^2
\overline{n}_k(\overline{n}_k+1),
\label{F2} \\
\tau_2^{-2}&=&{1\over 2}
\left(\sum_{\bf k}
{\partial
\epsilon_k(\overline{N}_0)
\over \partial \overline{N}_0}\overline{n}_k
\right)^2\langle \left(\delta {\partial N_0 \over 
\partial N}\right)^2\rangle \, . 
\label{F3} 
\end{eqnarray}
\noindent
In the term (\ref{F3})
there is the factor $\langle (\delta \partial  N_0 
/\partial N)^2\rangle$ which can be found from
Eq.(\ref{D5}).
The change of $N$ by 1 does
not affect the excitations accordingly to
the above discussion. Hence, the quasiparticle operators
in Eq.(\ref{D3}) should be taken as independent of $N$. Thus,
we find from (\ref{D5})

\begin{eqnarray}
\displaystyle {\partial N_0 \over \partial N}=1-
\sum_{\bf k}\left({\partial \over \partial N}(u^2_k)
\alpha^{\dagger}_{\bf k}\alpha_{\bf k}+
{\partial \over \partial N}(v^2_k)
\alpha_{\bf k}\alpha^{\dagger}_{\bf k}+
{\partial \over \partial N}(u_kv_k)
(\alpha^{\dagger}_{\bf k}\alpha^{\dagger}_{-\bf k}+
\alpha_{\bf k}\alpha_{-\bf k})\right)\, . 
\label{F7}
\end{eqnarray}
\noindent
This representation should be employed in Eq.(\ref{F3}).
As discussed in Ref.\cite{FLUC}, the main contribution
into $\langle \delta N_0^2\rangle$ comes from small momenta
where $u^2_k\approx v^2_k \sim \sqrt{\overline{N}_0}$. Thus,
Eq.(\ref{F7}) yields
$\partial N_0 /\partial \overline{N}_0\approx 
N_0/2\overline{N}_0$. Accordingly, 
$\langle (\delta \partial N_0 /\partial \overline{N}_0)^2
\rangle \approx \langle \delta N_0^2
\rangle /4\overline{N}_0^2$. Employing this as well as
the explicit form of $\langle \delta N_0^2\rangle$ 
for the box from Ref.\cite{FLUC}, we obtain for the 
dephasing rate (\ref{F1})

\begin{eqnarray}
\tau_d^{-1}=T\sqrt{{1.8\xi^{1/2}\over N}
+{6.0\xi^{2/3}\over N^{2/3}}\left({T\over T_c}\right)^3
},
\label{F8}
\end{eqnarray}
\noindent
where the limit $T\gg \mu$ is taken. 
The first term under the square root is due to
the ensemble fluctuations of the population
factors
(see Eq.(\ref{F2})).
This contribution discussed by the authors
in Ref.\cite{DAMOP} is effectively gained at the momenta
close to $k_c$. These fluctuations do not
practically contribute to the fluctuations
of the condensate population $N_0$ \cite{FLUC}. 
The second term (see Eq.(\ref{F3})) describes
the contribution of the ensemble fluctuations
of $N_0$ determined by the low momenta collective
modes \cite{FLUC}.
It can be seen, that the two terms under the square root
can be dominant in different regions of $N$ and $\xi$.

The result (\ref{F8}) is obtained for the box. It is clear
that, apart from the numerical coefficients, a similar expression
can be derived for the actual oscillator trap. 
Eq.(\ref{F8}) can be employed
for order of magnitude estimates for the conditions 
of the actual trap \cite{JILA}. 
At large $N$, the first term
behaves as $\sim N^{-4/5}$, and the second $\sim N^{-2/5}$,
provided $T/T_c=const$ and the condensate density
$n_0\sim N^{2/5}$ \cite{BAYM}. Taking values typical for
the experiment \cite{JILA}
$N_0=5\cdot 10^5$, $T\approx 50$nK, $a\approx 5\cdot 10^{-7}$cm,
$n_c\approx 10^{14}$cm$^{-3}$, we find $\xi=a^3n_c\approx 2\cdot
10^{-5}$, and $T_c\approx 3.3
\hbar^2n_0^{2/3}/m\approx 500nK$. For these
values, the second term under the square root
in Eq.(\ref{F8}) is much smaller than the first one due
to the temperature factor. In this case we find the dephasing
time $\tau_d\approx 1$s. If, however, temperature is increased
up to $T/T_c=0.5$,  
the second term becomes dominant.
In this case, the dephasing time becomes 
$\tau_d \leq 100$ms, which is within the experimental range
\cite{JILA}. 
In this estimate we assumed no shot noise due
to the uncertainty of the total value $N$.
If, however, the variance $\Delta N$ in (\ref{E22})
is large enough,
the total dephasing rate
$\sqrt{\tau_d^{-2}+(\tau_d^{(0)})^{-2}}$ can 
vary over a large range.

We note, however, that the results 
obtained above should not be directly  
applied to the experiment \cite{JILA} in which
the evolution of the relative phase of the two-component
condensate has been studied. In a following publication
we will modify the above analysis in order to apply it
to the experiment
\cite{JILA}.

\section{Discussion and summary}

In this work we have applied the concept of the
Orthogonality Catastrophe to a bosonic ensemble 
in order to treat temporal correlations of a
confined
Bose-Einstein condensate, consisting of a
finite and fixed (albeit large) number of bosons.
The occurrence
of the OC turns out to be a prerequisite to 
the irreversible
decay of such correlations. We stress that 
by saying "the temporal correlations of
the condensate" we mean the correlator
which is associated with
the evolution of the global phase of a single
condensate, or the relative phase between two condensates
which do not exchange particles.
The above analysis does not apply to the
decay of the normal excitations.

We have
shown that the OC occurs if some infinite
interacting bath contacts a finite condensate.
On the contrary, if no external bath is present,
the 
equilibrium evolution of the
condensate correlator can be treated as though
the normal component is not perturbed
at all by such an evolution. 
The correlations are dominated by the
reversible dephasing caused by the ensemble
averaging over the realizations of the
normal component.  
The effect of the {\it level repulsion}
is shown to be significant in suppressing
the irreversible decay at times much
longer than the dephasing time. 

In general,
relation between the three time scales -- the time
of reversible dephasing ~$\tau_d $~ induced by the ensemble
fluctuations of the chemical potential; 
the OC time ~$t_{OC}$~ (\ref{J2}); and the 
time ~$\Delta^{-1}$~, during which the effect of
the {\it level repulsion} takes place - determines
the nature of the condensate decoherence. In the isolated
trap, the case ~$\tau_d \ll t_{OC}$~ and ~$ t_{OC} \gg
\Delta^{-1}$~ is realized. Thus, the decoherence is
the reversible dephasing. In the presence of an external
bath, ~$t_{OC}\ll \Delta^{-1}$~, and the $strong$ OC
develops (the {\it level repulsion} is insignificant). 
In this case, the reversible dephasing can be
observed if ~$\tau_d < t_{OC}$~. If, however,
~$\tau_d > t_{OC}$~, the decoherence of 
the equilibrium condensate correlator is completely 
irreversible. Realization of either case depends
essentially on the nature of the external bath and its
interaction with the condensate. The {\it strong} 
OC with short ~$t_{OC}$~ deserves special attention
because in this case the nature of the low energy 
excitations may change dramatically \cite{RUK}.

Our results should be compared with those obtained
by Graham in Refs.\cite{GRAM1,GRAM2}. We note that
in the phenomenological approach of Refs.\cite{GRAM1,GRAM2},
the classical field $\langle \Psi (x,t)\rangle$ has
been treated as a dynamical variable coupled to the
reservoir of the normal excitations. A corresponding
evolution equation with the white noise due
to the normal component has been
postulated. The conclusion has been given
that a major reason for the decay of the 
field $\langle \Psi (x,t)\rangle$ is the
irreversibility developing at times longer than some
decay time containing a factor
larger than $\sqrt{N}$  (see, e.g., Eqs.(7.48)-
(7.50) of Ref.\cite{GRAM2}). 
In this regard we note that, the assumption that
$\langle \Psi (x,t)\rangle \neq 0$ is a classical variable
which satisfies classical equations of motion
inevitably violates the conservation of $N$ on the level
of an  uncertainty $\approx \sqrt{N}$. 
Thus, the classical Bogolubov treatment is accurate
within $1/\sqrt{N}$.
This does not
introduce any problem unless the long time limit
is considered (times $\sim \sqrt{N}$ and longer)
\cite{WALLS,KUK,PITCR,CASTIN}.
Thus, we consider the 
results of Refs.\cite{GRAM1,GRAM2}
as based on an unjustified use of the Gross-Pitaevskii
equation as well as of the Bogolubov
approximation in the domain where these approaches
should not be employed. 
In contrast, our treatment takes into account 
the conservation of $N$ explicitly, and the
Bogolubov approximation (in Sec.VI) was employed 
only for calculating the dephasing rate which itself 
has a prefactor $1/\sqrt{N}$, so that the corrections
$\sim 1/N$ due to the Bogolubov approximation
are insignificant. 
    
In future work we will extend our results
to the case of the specific situation 
realized by the JILA group \cite{JILA}. 
We will also discuss how 
the reversibility of the   
dephasing can be revealed
in the echo-type experiments \cite{DAMOP}. It is also
worth considering the possibility of creating
a situation in which a confined equilibrium 
condensate contacts some external bath inducing fast
irreversible decay of the condensate correlator \cite{RUK}.

\acknowledgements
One of the authors (A.K.) is grateful to 
E. Cornell for useful discussion 
of the possibilities of observing
the echo effect. A.K. is also thankful
to D.Schmeltzer
and A.Ruckenstein for reading the
manuscript during its preparation and for 
stimulating conversations. We also acknowledge
support from a CUNY collaborative grant.

\appendix
\section{Terms $ B_1(T)$ and $B_2(T)$ in Eqs.(\ref{N191}), (\ref{N192})} 
The terms $B_1(T)$ (\ref{N191}), and $ B_2(T)$ 
(\ref{N192}) can be written
as
\begin{eqnarray}
B_1(T)=-{2 n_0\over (4N)^2}\sum_m p_m \sum_{\bf k}
\widehat{\langle N,m|}(a^2_{\bf k}- a^{\dagger\,2}_{\bf k})
{\partial \over \partial n_0 }\widehat{|N,m\rangle }
, 
\label{A1}
\end{eqnarray}
\noindent
and
\begin{eqnarray}
B_2(T)=- {8n_0^2\over (4N)^2}\sum_m p_m 
 \widehat{\langle N,m|}
{\partial^2 \over \partial n^2_0} 
 \widehat{|N,m\rangle }
, 
\label{A2}
\end{eqnarray}
\noindent
because $\Phi'_m( \{\theta_{\bf k}\}, n_0)$ are the
{\it projected states} $\widehat{|N,m\rangle }$. In Eq.(\ref{A1})
we employed the standard representation 

\begin{eqnarray}
\theta'_{\bf k}= {a_{\bf k} + a^{\dagger}_{-\bf k} \over
\sqrt{2 z_k}},\quad {\partial \over \partial \theta'_{\bf k}}\,=
{(a_{\bf k} - a^{\dagger}_{ \bf k})
\sqrt{ z_k} \over
\sqrt{2}}, \quad z_k=\sqrt{{\hbar^2k^2n_r\over mg}} 
\label{A3}
\end{eqnarray}
\noindent
in terms of the phonon annihilation and creation operators $a_{\bf k}$
and $a_{\bf k}^{\dagger}$, respectively. These operators diagonalize
the free part (\ref{N14}) of the $projected$
Hamiltonian (\ref{N13})-(\ref{N15}), 
so that $H'_0=\sum_{\bf k}\epsilon_k
a_{\bf}^{\dagger} a_{\bf k}$, with the bare spectrum defined as
$\epsilon_k=c_0k$, $c_0=\sqrt{\hbar^2n_rg/m}$.

The matrix elements $\widehat{\langle N,m'|}
{\partial \over \partial n_0} 
 \widehat{|N,m\rangle }$ can be found by
differentiating the relation $H(N) \widehat{|N,m\rangle }=E_m(N)
\widehat{|N,m\rangle }$ with respect to $n_0$ (note that $n_0=N/V$ 
is the only combination which contains the dependence on $N$). Then,
we find 

\begin{eqnarray}
\widehat{\langle N,m'|}
{\partial \over \partial n_0} 
 \widehat{|N,m\rangle }=-{3\over 4n_0}{(H'_{int})_{m'm}\over 
E_m(N)-E_{m'}(N)}
\label{A4}
\end{eqnarray}
\noindent
for $m\neq m'$ and zero otherwise. In here we employed an
explicit form (\ref{N15}). We note that (\ref{A4}) is
exact. Then, (\ref{A1}) can be represented as

\begin{eqnarray}
B_1(T)={3\over 2(4N)^2}\,\sum_{\bf k}
\sum_{m,m'} p_m (a^2_{\bf k}- a^{\dagger\,2}_{\bf k})_{mm'}
{(H'_{int})_{m'm}\over 
E_m(N)-E_{m'}(N)}.
\label{A5}
\end{eqnarray}
\noindent
The expression (\ref{A5}) can be evaluated by employing
the perturbation theory with respect to
$H'_{int}$ (\ref{N15}). We note that in the lowest order,
$B_1(T)=0$ because the zeroth order intermediate states
$\widehat{|N,m'\rangle }$ differ from the
state $\widehat{|N,m\rangle }$ by two phonons, while  
$(H'_{int})_{m'm}$ (\ref{N15}) links those states which differ by either
one or three phonons. The higher order terms do make
$B_1(T)\neq 0$. However, the dominator $ E_m(N)-E_{m'}(N)$ in
this expression may lead only to the first order pole which
does not lead to the divergence $\sim 1/\omega^2$ in the
overlap (\ref{I2}). Thus, the term $B_1(T)\sim 1/N$.

The term $B_2(T)$ in (\ref{A2}) can be rewritten by
employing the relation $ \widehat{\langle N,m|}
{\partial^2 \over \partial n^2_0}
 \widehat{|N,m\rangle }=- ({\partial \over \partial n_0}
\widehat{\langle N,m|})({\partial \over \partial n_0}
\widehat{|N,m\rangle })$ as well as
Eq.(\ref{A4}). Finally, we find

\begin{eqnarray}
B_2(T)={1\over 2}\left({3\over 4N}\right)^2
\sum_{m,m'} p_m 
{|(H'_{int})_{m'm}|^2\over 
(E_m(N)-E_{m'}(N))^2}.
\label{A6}
\end{eqnarray}
\noindent
This expression coincides with ${1\over 2}\int\, d\omega J(\omega)
/\omega^2$ where $J(\omega)$ is given by Eq.(\ref{N22}). 

\section{Level repulsion and absence of the OC in an isolated trap }

We evaluate the typical matrix
element ~$\Delta \approx |(H'_{int})^{(0)}_{mn}|$,
where the index ~$^{(0)}$~ indicates that
the matrix elements are taken in zeroth
order with respect to $H'_{int}$ (\ref{N15}).
At high $T$ one can estimate
~$\theta_k \sim \sqrt{mT/\hbar^2k^2n_r}$~ 
and ~$\partial .../\partial
\theta_k \sim \sqrt{T/g}$~ from
(\ref{N14}). Accordingly,
we arrive at
the estimate of $\Delta$ from Eqs.
(\ref{N13})-(\ref{N15}) as  

\begin{eqnarray}
\Delta  \approx {T\over \sqrt{N}}
\left({T\over \mu}\right)^{1/2}.   
\label{J8}
\end{eqnarray}
\noindent
This value sets up an energy scale below which
the {\it level repulsion} becomes significant.
Here the factor $1/\sqrt{N}$ indicates that
no the {\it strong OC} occurs. Indeed, as discussed
in Sec.IVB, the necessary condition for the
{\it strong OC} is that the OC time $t_{OC}$ 
(\ref{N210})
(calculated by means of the perturbation expansion)
is
shorter than $\Delta^{-1}$. Eq.(\ref{N210})
yields ~$t_{OC}\sim N$, implying that for large
enough $N$, the opposite limit holds 
~$ t_{OC}\gg \Delta^{-1}$. Hence, the OC in the isolated
trap may occur only as the {\it weak OC} (see Sec.IVC).

The relevant frequency scale, at which the effect
of the {\it level repulsion} dominates, is
 given by

\begin{eqnarray}
\omega \leq \Delta .  
\label{J80}
\end{eqnarray}
\noindent
At such frequencies
the {\it level repulsion} effect strongly modifies
the spectral weight (\ref{N22}) with respect to the
perturbation expansion.
As discussed in Sec.IVC,  
a strong renormalization of the matrix elements
takes place. This effect is non-perturbative.
This can be
understood within the effective two-states model, 
in which 
a pair of the
degenerate states  ~$\widehat{|m,N\rangle}^{(0)}$~
and ~$\widehat{|n,N\rangle}^{(0)}$~ defined with respect to
the free Hamiltonian $H'_0$ (\ref{N14}) is considered
independently from the rest of the states. 

In order to justify this two-state approximation we note 
that 
the degeneracy contributing to the spectral weight
$J(\omega)$ (\ref{N22}) at $\omega \to 0$ in the lowest
order of the perturbation expansion 
is associated with the processes
of the decay and the recombination of two phonons. 
Given an initial phonon with some momentum $\bf k$, the final
states contain two phonons with the momenta $\bf p$ and 
$\bf q$, which obey the conservation
condition ${\bf k}={\bf p} + {\bf q}$.
The subspace of the final states, which may become
degenerate with the initial state, is characterized
by different ${\bf p}$ and ${\bf q}$.
Only those final states, which are close enough in 
energy to the initial state, participate practically
in the 
renormalization caused by the {\it level repulsion}.
A natural criterion for the closeness is given by the 
typical value of the matrix element $\Delta$ (\ref{J8}). 
Let us assume that some pair ${\bf p}$ and ${\bf q}$
satisfies this condition. Then, 
in a finite system, another closest final 
state will be characterized by the momenta differed
by the amount $\sim 1/L\to 0$, where $L$ is a typical
size of the system. Accordingly, their energies 
should also differ by some amount $ \sim 1/L\to 0$.
The value of $\Delta$ (\ref{J8}) is scaled, however,
as $\sim 1/L^{3/2}\ll 1/L$. Thus, for large enough system,
such a closest state turns out to be out 
of the
range of the {\it level repulsion}. These considerations
justify that only pairs of the degenerate states participate
practically in the {\it level repulsion}. 

Then,
the matrix element ~$(H'_{int})^{(0)}_{mn}$~, which is 
defined with respect to the bare states
~$\widehat{|m,N\rangle}^{(0)}$~  
and ~$\widehat{|n,N\rangle}^{(0)}$, mixes these states
(as long as ~$|\omega^{(0)}_{mn}|\leq 
\Delta$). 
The new states
~$\widehat{|m,N\rangle}$~ and ~$\widehat{|n,N\rangle}$~
are the linear combinations of the bare states
~$\widehat{|m,N\rangle}^{(0)}$~
and ~$\widehat{|n,N\rangle}^{(0)}$. Then, 
the standard treatment of the two-level system
close to the degeneracy shows that the renormalized
matrix element  ~$(H'_{int})_{mn}$~  
and the energy difference $\omega_{mn}$ become 

\begin{eqnarray}
(H'_{int})_{mn}={|\omega^{(0)}_{mn}|\over
|\omega_{mn}|}(H'_{int})^{(0)}_{mn},\quad
\omega_{mn}=\pm
\sqrt{|(H'_{int})^{(0)}_{mn}|^2 + 
(\omega^{(0)}_{mn})^2}.
\label{L1}
\end{eqnarray}
\noindent
Thus,
~$(H'_{int})_{mn}\sim \omega^{(0)}_{mn}\to 0$~
for ~$|\omega^{(0)}_{mn}|\ll \Delta$. 
This feature alone leads to a strong
suppression of ~$J(\omega)$~ (\ref{N22}) in the range
(\ref{J80}).

Let us estimate the mean overlap ~$\overline{\chi}$
(\ref{N21}) where the $J(\omega)$ is given by
Eq.(\ref{N22}).
We introduce a distribution function of
the bare matrix elements ~$|(H_{int})_{mn}^{(0)}|$~
and of the excitation energies ~$|\omega_{mn}^{(0)}|$~:

\begin{eqnarray}
P(M,\omega_0)=\sum_{m,n\neq m}p_m
\delta(M-|(H_{int})_{mn}^{(0)}|)\delta(\omega_0-
|\omega_{mn}^{(0)}|),
\label{B1}
\end{eqnarray}
\noindent
so that the spectral weight (\ref{N22}) can be represented
as

\begin{eqnarray}
J(\omega)= \left({3\over 4N}\right)^2
\int_0^{\infty}\,dM\int_0^{\infty}\,d\omega_0\,
P(M,\omega_0){\omega_0^2\over \omega^2}M^2
\delta(\omega-\sqrt{M^2 +\omega_0^2}) ,
\label{B2}
\end{eqnarray}
\noindent
where we have employed Eq.(\ref{L1}). 

We assume that, if no {\it level repulsion} is
taken into account (that is,
the matrix elements as well as
the excitation energies are taken unrenormalized),
the spectral weight (\ref{N22})
can be well taken as a constant.
This would imply the OC. A 
simplest form of $P(M,\omega_0)$, which will
produce such a spectral weight in Eq.(\ref{B2}),
if the {\it level repulsion} were ignored, is 

\begin{eqnarray}
P(M,\omega_0)=P_0\exp (-{M\over \Delta}
-{\omega_0 \over \omega_c}),
\label{B3}
\end{eqnarray}
\noindent
where $P_0$ is some normalization constant
which can be found from the sum rule

\begin{eqnarray}
\int\,d\omega J(\omega)=(3/4N)^2\langle 
(H'_{int})^2\rangle
\label{B5}
\end{eqnarray}
\noindent
following from
Eq.(\ref{N22}). The form (\ref{B3}) indicates that 
the distribution of the matrix elements and the excitation
frequencies is uniform, as long as the matrix elements are
less than  some
typical element ~$\Delta$~ (\ref{J8}) and the excitation
frequencies are below the high energy cut-off ~$\omega_c$. 
Employing (\ref{B3}), the
unrenormalized spectral weight acquires 
the form

\begin{eqnarray}
\displaystyle 
J^{(0)}(\omega)= \left({3\over 4N}\right)^2P_0
\int_0^{\infty}\,dM\int_0^{\infty}\,d\omega_0\,
{\rm e}^{-{M\over \Delta}
-{\omega_0 \over \omega_c}}\,\,M^2
\delta(\omega-\omega_0)=2 
\left({3\over 4N}\right)^2\Delta^3P_0
{\rm e}^{-{\omega \over \omega_c}}.
\label{B4}
\end{eqnarray}
\noindent
It satisfies $J(0)\neq 0$. In order to obtain
$P_0$, it is enough to employ this
unrenormalized form in Eq.(\ref{B5}), 
because the renormalization
(\ref{B2}) becomes important only for 
~$\omega \leq \Delta$~, where ~$\Delta$~ in
Eq.(\ref{J8}) is small
as $\sim 1/\sqrt{N}$. Then, a substitution
of (\ref{B4}) into (\ref{B5}) yields

\begin{eqnarray}
P_0= {1\over 2\omega_c\Delta^3}\langle 
(H'_{int})^2\rangle.
\label{B6}
\end{eqnarray}
\noindent
This can now be employed in Eq.(\ref{B3}),
and then in Eq.(\ref{B2}). This yields
the renormalized $J(\omega)$ (\ref{B2}) 
as 

\begin{eqnarray}
\displaystyle 
J(\omega)= \left({3\over 4N}\right)^2
{\langle(H'_{int})^2\rangle
\over 2\omega_c}\left({\omega\over \Delta}\right)^3
\int^1_0 dxx^2\sqrt{1-x^2}{\rm e}^{-\omega x/\Delta},
\label{B40}
\end{eqnarray}
\noindent
where the condition ~$\omega \ll \omega_c$
is taken into account.
Thus, ~$J(\omega)\sim \omega^3/N$~
for ~$\omega \ll \Delta$~, and
~$J(\omega)\approx const\sim 1/N$~ for
~$\Delta \ll \omega \ll \omega_c$.
Such a behavior implies that the OC is eliminated due to
the {\it level repulsion}. This can be seen 
directly after employing Eq.(\ref{B40})
in Eq.(\ref{N21}) for the mean overlap. 
We find

\begin{eqnarray}
\ln \overline{\chi} =-{9\pi \over 2^8}
{\langle (H'_{int})^2\rangle, 
\over N^2 \omega_c \Delta}
\label{L2}
\end{eqnarray}
\noindent
A quick glance on Eq.(\ref{L2}) shows that $\ln 
\overline{\chi}\sim -1/\sqrt{N}\to 0$. Indeed,
$ \langle (H'_{int})^2\rangle $ is the mean square
interaction energy $\sim N$. The value of $\Delta$
is $\sim 1/\sqrt{N}$ (\ref{J8}). Thus, due to the
$N^2$ in the denominator, the total factor $1/\sqrt{N}$
follows. Accordingly, $\overline{\chi}=1+o(1/\sqrt{N})$
implying no the OC.


\begin{references}
\bibitem{REM1}
For example, consider a situation 
when $N$ is fixed, so that
no broken gauge symmetry can strictly be introduced.
\bibitem{REM2}
such a time at zero temperature is approximately determined
as $\sim (N/\delta N E_{in})$, where $\delta N$ stands for
the variance of $N$ in the coherent state, and 
$ E_{in}$ is a typical interaction energy
per boson \cite{WALLS}.
\bibitem{REM5}
Such $1/N$ expansion should not be identified with 
the $1/N$ expansion employed in Ref.\cite{CASTIN}.
\bibitem{RUK}
As pointed out by A. Ruckenstein ,
it would be interesting to find an example of
the system 
where $ t_{OC}$ is short enough
to be comparable with typical inverse
frequencies of the elementary excitations.
In this case, the 
low energy elementary excitations
become ill defined even in the case of the weakly interacting
bosons.
\end{references}
\end{document}